\documentclass{aastex}
\usepackage{emulateapj5}
\begin{document}

\newcommand{\hi}{$h^{-1}$~}
\newcommand{\kms}{~km~s$^{-1}$~}
\newcommand{\rh}{$r_{1/2}$}
\newcommand{\logh}{+5log$h$}
\newcommand{\eg}{$e.g.$}

\title{CL~1358+62:  Characterizing the Physical Properties of Cluster 
Galaxies at $z=0.33$}

\author{Kim-Vy H. Tran\footnotemark[1]}
\footnotetext[1]{Current addresss:  Institute for
  Astronomy, ETH H\"onggerberg HPF G4.2, CH-8093 Z\"urich, Switzerland}
\affil{Department of Astronomy \& Astrophysics, University of
California, Santa Cruz, CA 95064}
\email{vy@phys.ethz.ch}
\setcounter{footnote}{1}

\author{Luc Simard}
\affil{National Research Council of Canada, Herzberg Institue of
Astrophysics, 5071 West Saanich Road, Victoria, BC V9E 2E7, Canada}
\email{Luc.Simard@nrc.ca}

\author{Garth Illingworth}
\affil{University of California Observatories/Lick Observatory,
University of California, Santa Cruz, CA 95064}
\email{gdi@ucolick.org}

\author{Marijn Franx}
\affil{Leiden Observatory, P.O. Box 9513, 2300 RA Leiden, The
Netherlands}
\email{franx@strw.leidenuniv.nl}

\begin{abstract}

We examine the physical properties of 173 cluster members in
CL~1358+62 ($z=0.3283$) from $HST$ WFPC2 imaging taken in the F606W
($\sim$rest-frame $B$) and F814W ($\sim$rest-frame $V$) filters over a
$2.2\times2.2$ Mpc$^2$ field ($H_0=100$ km~s$^{-1}$~Mpc$^{-1}$,
$\Omega_M=0.3$, $\Omega_{\Lambda}=0.7$). Structural parameters are
measured for each galaxy by fitting a PSF-convolved, two component
model to their 2D surface brightness distribution.  We examine
bulge+disk models using three different bulge profiles (de Vaucouleurs,
S\'ersic, and exponential), and rigorously test the robustness of our
results by analyzing several thousand artificial galaxies in the same
manner as the cluster data.  The measured physical properties from the
best-fit profile of the cluster galaxies are combined with ground-based
spectroscopy to test for correlations between morphological
characteristics, current star formation, total galaxy colors, and
cluster substructure.  We find that: (1) Bulge-to-total ratio
[$(B/T)_{deV}$] and Hubble type ($-5\leq T\leq8$) are strongly
correlated (99\% confidence), but the scatter is large and early-type
spirals are not reliably distinguished from ellipticals and S0's based
on $(B/T)_{deV}$.  (2) From comparison of their physical properties,
the low luminosity ($-17.3\geq MB_z-5\log h\geq-19.3$) ellipticals
in our sample are likely to be face-on S0 galaxies.  (3) High galaxy
asymmetry and strong [OII]$\lambda3727$ emission are strongly
correlated for disk-dominated members [$(B/T)_{deV}<0.4$]. (4) There
exists a small population ($\sim5$\%) of bulge-dominated members whose
significant [OII]$\lambda3727$ emission ($<-5$\AA) suggest they harbor
active galactic nuclei.  (5) At these redshifts, determining the
correct S\'ersic index $n$ can be highly unreliable.

\end{abstract}



\section{Introduction}

How galaxies evolve as a function of their environment remains a
fundamental question in astronomy.  In particular, the environment of
rich clusters provides a unique laboratory for studying how the
physical properties of galaxies are related to varying local density,
interactions with other galaxies, and exposure to a hot intracluster
medium.  One deeply ingrained and useful approach to studying galaxy
evolution is characterizing morphological properties by visually
typing galaxies using the Hubble system \citep{hubble:26}.  As we move
into the era of galaxy surveys with thousands, if not millions, of
objects, however, there is a need for a quantitative, uniform, and
{\it reproducible} method of cataloging these visual properties.  

A promising solution to this challenge is to measure physical
properties, \eg~bulge and disk scale lengths, bulge-to-total ratios,
disk inclination, bulge ellipticity, etc., directly from 2D surface
brightness distributions.  Here, parametric models are convolved with
the point spread function and then compared to the galaxy's image to
determine the best fit model
\citep{schade:95,dejong:96,marleau:98,simard:99,tran:01,trujillo:01,simard:02,labarbera:02,macarthur:02}.
Not only are 2D fits a natural way to weight the information contained
in each pixel, they are crucial for detecting the presence of bars,
HII regions, tidal tails, arms, shells, etc., $i.e$ structures not
easily identified in traditional 1D profiles.  By keeping the 2D
information, structural parameters for a large sample of galaxies can
be derived and then compared to the galaxies' spectral properties and
environmental conditions.  A lingering concern, however, is assessing
how meaningful the structural parameters are.  In the well-accepted
framework of Hubble classifications, how can bulge-to-total ratios,
scale lengths, galaxy asymmetry, etc. aid our understanding of galaxy
evolution?

In Hubble's original system \citep{hubble:26}, one of the main
qualitative classification criteria is bulge-to-disk ratio.  As
bulge+disk models measure a basic extension of this parameter
(bulge-to-total fraction; $B/T$), it should be possible to correlate
bulge fraction with Hubble type.  The bulge fraction varies, however,
depending on the type of bulge profile used, $i.e.$ classical de
Vaucouleurs $r^{1/4}$ law \citep{devaucouleurs:48}, the generalization
by S\'ersic to $r^{1/n}$ \citep{sersic:68}, or pure exponential
\citep{dejong:96}.  A fundamental question is which profile correlates
best to Hubble classifications for galaxies spanning the range of
morphological types?

Once the appropriate profile is determined, the measured structural
parameters provide a wealth of information with which to study
galaxies.  In addition to confirming the morphology-density relation in
intermediate redshift clusters \citep{dressler:97}, recent studies find
a population of late-type/disk-dominated members that lack strong
emission \citep{poggianti:99,balogh:02}, i.e. their star formation has
been quenched.  This result suggests that spectral and morphological
evolution in the cluster environment may be decoupled and raises the
question of how strongly ongoing star formation is correlated with
physical characteristics.  By comparing spectral features to
quantitative measures of galaxy morphology from our extensive sample of
CL~1358+62 \citep[$z=0.3283$;][hereafter F98]{fisher:98}, we can
address this issue as well as test for correlations between morphology
and cluster substructure \citep[\eg][]{hutchings:02}.

A direct test of the link between star formation and morphology is to
compare [OII] emission to residuals from the galaxy fits.
\citet{couch:01} find that spiral structure still can be observed in
$z\sim0.3$ cluster galaxies long after star formation has ceased,
$i.e.$ signatures of past events may be long-lived.  If 1)
morphological disturbances associated with star formation continue to
exist long after the star forming episode ends, and 2) late-types are
``strangled'' when they enter the cluster environment \cite[$R>2$\hi
Mpc]{balogh:98}, then we should observe a population of predominantly
late-type galaxies in CL~1358+62 with high asymmetry but no detectable
[OII]$\lambda3727$ emission.  By comparing [OII]$\lambda$3727 to
bulge-to-total ratio and galaxy asymmetry, we can isolate these
galaxies.

Another interesting question that can be addressed using 2D fits is
whether low luminosity ellipticals and S0's are essentially the same
type of galaxy.  It has been argued that $M_B\geq-22$ ellipticals and
S0's have different formation/merger histories than the most luminous
early-types.  \citet{jorgensen:94} suggest that these fainter
ellipticals and S0's actually form a common parent galaxy population,
and that viewing angle plays a strong part in how these galaxies are
typed \cite[see also]{rix:90,kormendy:96}.  Their similarity in colors
support this argument \citep{sandage:78}.  As the isophotal shapes of
early-type galaxies hold clues to their formation history
\citep{nieto:89,naab:99}, determining if bulge-dominated systems must
be separated into two classes, $i.e.$ ``boxy'' bright ellipticals and
``disky'' fainter ones \citep{kormendy:96}, is integral to
understanding galaxy evolution across the range of luminosity
($\sim$mass).  To address this question, we compare the bulge
fractions, disk inclinations, bulge ellipticities, and velocity
dispersions of visually classified ellipticals and S0's in CL~1358+62
\cite[hereafter FFvD00]{fabricant:00}.

In this paper, we characterize the physical properties of 173
confirmed members in CL~1358+62 (F98) using GIM2D
\citep{marleau:98,simard:99,tran:01,simard:02}, an automated program
that fits a PSF-convolved 2D surface brightness model to an image of
each galaxy and searches $\chi^2$ space for the best fit.  The images
are from $8'\times8'$ $HST$ WFPC2 mosaics taken in the F606W and F814W
filters ($\sim$ rest-frame $B$ and $V$).  The members span the range
of Hubble type ($-5\leq T\leq8$) and have apparent F814W magnitudes
between $17.4$ (BCG) and $23.2$ ($-20.8\geq MB_z-5\log
h\geq-15.9$).  By fitting three different bulge+disk profiles to
the sample, we determine the profile that is correlated best with the
published visual classifications.  We test for correlations between
structural parameters, current star formation as defined by
[OII]$\lambda3727$ emission, internal velocity dispersions, and local
cluster structure.  We also investigate the relationship between
elliptical and S0 galaxies, and examine the properties of E+A
galaxies.

To determine how robust our results are, we simulate thousands of
artificial de Vaucouleurs bulge+exponential disk galaxies and fit
surface brightness profiles to them in the same manner as with the
cluster sample; note we assume galaxies can be well-described by a two
component profile.  The artificial galaxy catalog covers the same range
in structural parameter space as the cluster members, and from these
simulations we quantify systematic and random errors associated with
bulge fraction, half-light radius, bulge \& disk scale lengths, disk
inclination, and bulge ellipticity.  We also test our ability to
measure S\'ersic and exponential bulge components by fitting artificial
galaxies with $r^{1/n}$ ($0.2\leq n\leq6$) bulge+exponential disk and
double exponential profiles.

The sections of this paper are organized as follows: The data and
fitting technique are described in \S2.  In \S3, we present the models
and structural parameters determined by our surface brightness fitting
program for the cluster sample.  Here we employ ``sanity checks'' by
comparing the measured bulge fraction to visual morphologies, local
cluster density, and total galaxy colors.  Correlations between bulge
fraction, total residual, galaxy asymmetry, and current star formation
also are examined in this section.  We discuss the physical properties
of elliptical, S0, and E+A galaxies in \S4.  In \S5, we test the
robustness of our measurements by utilizing results from thousands of
artificial galaxies fits.  Our conclusions are summarized in \S6.
Here we use $H_0=100 h$\kms Mpc$^{-1}$, $\Omega_M=0.3$, and
$\Omega_{\Lambda}=0.7$.

\section{Data and Analysis}

\subsection{Imaging}

We use $HST$ WFPC2 imaging of CL~1358+62 that is part of an extensive
project to study the evolution of intermediate redshift cluster
galaxies
\citep{vandokkum:98b,vandokkum:98a,tran:99,vandokkum:99,vandokkum:00,kelson:00a,kelson:00b,kelson:00c,kelson:01,vandokkum:01}.
The $8'\times8'$ mosiac ($R\sim1.1$\hi Mpc) is centered on the
brightest cluster galaxy (BCG) and is comprised of 12 separate
pointings in the F606W and F814W filters. Three exposures, each 1200
seconds, were taken in each filter for each pointing.  The two chosen
filters, F606W and F814W, are similar to rest-frame $B$ and $V$ for
CL~1358+62.  The images were reduced at the Space Telescope Science
Institute (STScI) with the usual pipeline procedure.  Details of the
complex sky subtraction and cosmic ray removal are explained fully in
\citet[hereafter vD98]{vandokkum:98a}.  Following vD98, we transform
from observed F606W and F814W magnitudes to rest-frame $B$ and $V$
using

\begin{equation}
B_z=F814W+1.021(F606W-F814W)+0.524
\end{equation}
\begin{equation}
V_z=F814W+0.204(F606W-F814W)+0.652
\end{equation}          

\noindent with a distance modulus of $40.03$; we also correct for passive 
evolution \citep{vandokkum:98b}.

\subsection{Spectroscopy}

The large spectroscopic sample obtained for the CL~1358+62 field was
taken at the Multiple Mirror Telescope and the William Herschel
Telescope.  A detailed description of the reduction and analysis of
the spectroscopic sample are described in \citet{fabricant:91} and
F98.

Twenty slit-masks were designed to cover a $10'\times11'$ field;
targets were chosen based on their $R$ magnitude from imaging taken at
the Whipple Observatory 1.2m telescope.  The spectral resolution was
$\sim13$\AA~at the WHT and $\sim20$\AA~at the MMT.  Of the 387
redshifts measured, 232 are cluster members.  Defining galaxies with
$0.31461<z<0.34201$ as cluster members, CL~1358+62's mean redshift and
velocity dispersion are $0.3283\pm0.0003$ and $1027^{+51}_{-45}$\kms
respectively (F98).

From the redshift catalog, 192 cluster members fall on the
$8'\times8'$ $HST$ WFPC2 mosaic.  The completeness within the $HST$
mosaic is $>90\%$ to $R=21$ ($m_{814}\sim20.2$), decreasing to
$\sim30\%$ at $R=22$.  The incompleteness is not due to an inability
to measure redshifts at these magnitudes but by the limited number of
galaxies observed.  The redshift success rate is not strongly color
dependent for $R<23.5$ (F98).

Of the 192 galaxies that are on the $HST$ mosaic, we fit galaxy models
only to 173 as 19 fall too close to a chip edge for proper analysis of
the surface brightness distribution.  The 173 galaxies range in
absolute $B_z$ magnitude from $-20.8$\logh~mags (H375; BCG) to
$-15.9$\logh~mags (I1829).  We include the published
[OII]$\lambda3727$ \AA~line strength measurements (F98) and internal
velocity dispersions for a subset \citep{kelson:00b} in our analysis.

\subsection{Structural Measurements}

\subsubsection{The Surface Brightness Models}

We use the GIM2D package
\citep{marleau:98,simard:99,tran:01,simard:02} to find the best-fit
PSF-convolved, 2D bulge+disk models to the surface brightness profiles
of the cluster members.  The program has a maximum of 12 fitting
parameters: the flux ($F_{total}$) in the model integrated to
$r=\infty$; the bulge/total luminosity $B/T \equiv
F_{bulge}/F_{total}$; the semi-major axis effective radius of the
bulge $r_e$; the bulge ellipticity $e\equiv1-b/a$ where $a$ and $b$
are the bulge semi-major and semi-minor axes respectively; the bulge
position angle $\phi_b$; the semi-major axis exponential disk scale
length $r_d$; the inclination of the disk $i$ ($i \equiv 0$ for
face-on); the disk position angle $\phi_d$; the subpixel $dx$ and $dy$
offsets of the galaxy's center; the residual background level $db$;
and the S\'{e}rsic index $n$ of the bulge.  Both $\phi_b$ and $\phi_d$
are measured clockwise from the positive $y$-axis of the image.  The
best-fit parameters and their confidence intervals are determined
using the Metropolis algorithm \citep{metropolis:53,saha:94} which
uses the $\chi^2_{\nu}$ test to determine the region of maximum
likelihood in the multi-parameter space.

The bulge profile is defined as
\begin{equation}
\Sigma(r) = \Sigma_e~exp\left\{ -k[(r/r_e)^{1/n} - 1] \right\}
\label{1358bulge-prof}
\end{equation}
\noindent where $\Sigma(r)$ is the surface brightness at $r$ along the 
semi-major axis, and $\Sigma_e$ is the effective surface brightness.
This bulge profile is also known as the S\'{e}rsic profile
\citep{sersic:68}.  The parameter $k$ is equal to $(1.9992n -
0.3271)$, a value that defines $r_e$ to be the projected radius
enclosing half of the light in the bulge component \citep{capaccioli:89}.
The classical de Vaucouleurs profile is a special case of
Equation~\ref{1358bulge-prof} with $n = 4$.

The disk profile is defined as
\begin{equation}
\Sigma(r) = \Sigma_0~exp(-r/r_d)
\label{1358disk-prof}
\end{equation}
\noindent where $\Sigma_0$ is the (face-on) central surface
brightness.  We note (as do \citealt{simard:99,simard:02}) that the
bulge/disk nomenclature adopted here to describe our surface
brightness models may not reflect the internal kinematics of its
components.  A ``bulge'' may not be a centralized, dynamically hot
spheroid but could be a central starburst.  Similarly, a ``disk'' may
not necessarily be a cold, co-rotating population.  For example,
dynamically ``hot'' systems such as faint dwarf ellipticals are best
fit by exponential disks \citep{binggeli:91,ryden:99}.

Before a 2D model to the surface brightness of a given galaxy can be
fitted, the galaxy's isophotal area and an appropriate point spread
function (PSF) must be determined.  To define the isophotal area, we
use the galaxy photometry package SExtractor V2.0 \citep{bertin:96}
with a detection threshold of $\mu_{814}=24.4$~mags/$\Box''$
(equivalent to $\sim1\sigma$ of the sky noise) and a minimum deblending
contrast parameter of 0.01.  As we fit galaxies in two filters (F606W
and F814W), we use the isophotal area defined in the (redder) F814W
image.  To generate a PSF for each galaxy in each filter, we use
TinyTim V4.4 \citep{krist:93}.  As the PSF changes across each WF chip,
a PSF model is generated every 50 pixels and the nearest one to the
galaxy is chosen for the GIM2D analysis.  Both the PSF and galaxy
models are subsampled by a factor of five as WF image are undersampled
and the pixelization can affect the shape of small galaxies such as
those in our cluster sample.  The centering of the galaxy is also
improved by subsampling the data; this point can be critical for
determination of the galaxy residuals \citep{conselice:00}.

By fitting models to the surface brightness distribution of these
galaxies, we measure the structural properties $n$, $B/T$, $r_e$,
$r_d$, $\phi_b$, $\phi_d$, $i$, and the half-light radius \rh.  The
semi-major axis half-light radius is computed by integrating the sum
of Equations~\ref{1358bulge-prof} and~\ref{1358disk-prof} to
$r=\infty$; note that as the models include bulge ellipticity and disk
inclination, circular symmetry is {\it not} assumed.  The galaxy's
$(x,y)$ center is also determined from the best fit model.

The asymmetric image residual flux is quantified by the asymmetry
index $R_{A}$
\citep{schade:95}, defined as
\begin{eqnarray}
R_A & = & (R_A)_{raw} - (R_A)_{bkg} \nonumber \\
& = & {{\displaystyle\sum_{i,j} {1\over{2}}| R_{ij} - 
R_{ij}^{180}|} \over{\displaystyle\sum_{i,j} I_{ij}}} - {{\displaystyle\sum_{i,j
} 
{1\over{2}}| B_{ij} - 
B_{ij}^{180}|} \over{\displaystyle\sum_{i,j} I_{ij}}}
\label{1358ra-index}
\end{eqnarray}
where $R_{ij}$ is the flux at $(i,j)$ in the residual image,
$R_{ij}^{180}$ is the flux in the residual image rotated by
$180^{\circ}$, and $I_{ij}$ is the flux in the original image.
Following \citet{marleau:98}, $R_A$ is measured within $r=2$\rh.  The
second term $(R_A)_{bkg}$ in Equation~\ref{1358ra-index} is a
statistical correction for background noise fluctuations.  Since
$(R_A)_{raw}$ involves taking absolute values of pixel fluxes, it will
yield a positive signal even in the sole presence of noise.  The
background correction, $(R_A)_{bkg}$, is computed over pixels flagged
as background pixels by SExtractor.  The $B_{ij}$'s are background
pixel values in the residual image, and the $B_{ij}^{180}$'s are
background pixel values in the residual image rotated by
$180^{\circ}$.  The background correction is computed over a
background pixel area equal to the pixel area over which $(R_A)_{raw}$
is computed.  Given the statistical nature of $(R_A)_{bkg}$, there
will be cases when a galaxy is faint enough compared to its background
noise that $R_A$ may take on small negative values in exactly the way
as the difference of two values of $(R_A)_{bkg}$ computed from
different regions of the sky may be negative.

In addition to $R_{A}$, we measure the total residual fraction of
light $R_T$ by taking the pixels assigned to the galaxy by SExtractor
and creating a mask that is applied to the original and model images
\citep{tran:01}.  For a positive-definite residual
fraction, the model is subtracted from the original, and the absolute
value of the difference at each pixel over the isophotal area is
summed.  To account for the sky, the same number of sky pixels as
galaxy pixels are summed in the same manner and subtracted from the
total galaxy residual.  We use the total residual fraction of light as
a gauge of the model's goodness of fit but note that, for fainter
galaxies, the error in $R_T$ is dominated by the error in the sky
flux.

\section{Physical Properties}

\subsection{What Type of Profile?}

\begin{figure}
\epsscale{0.8}
\plotone{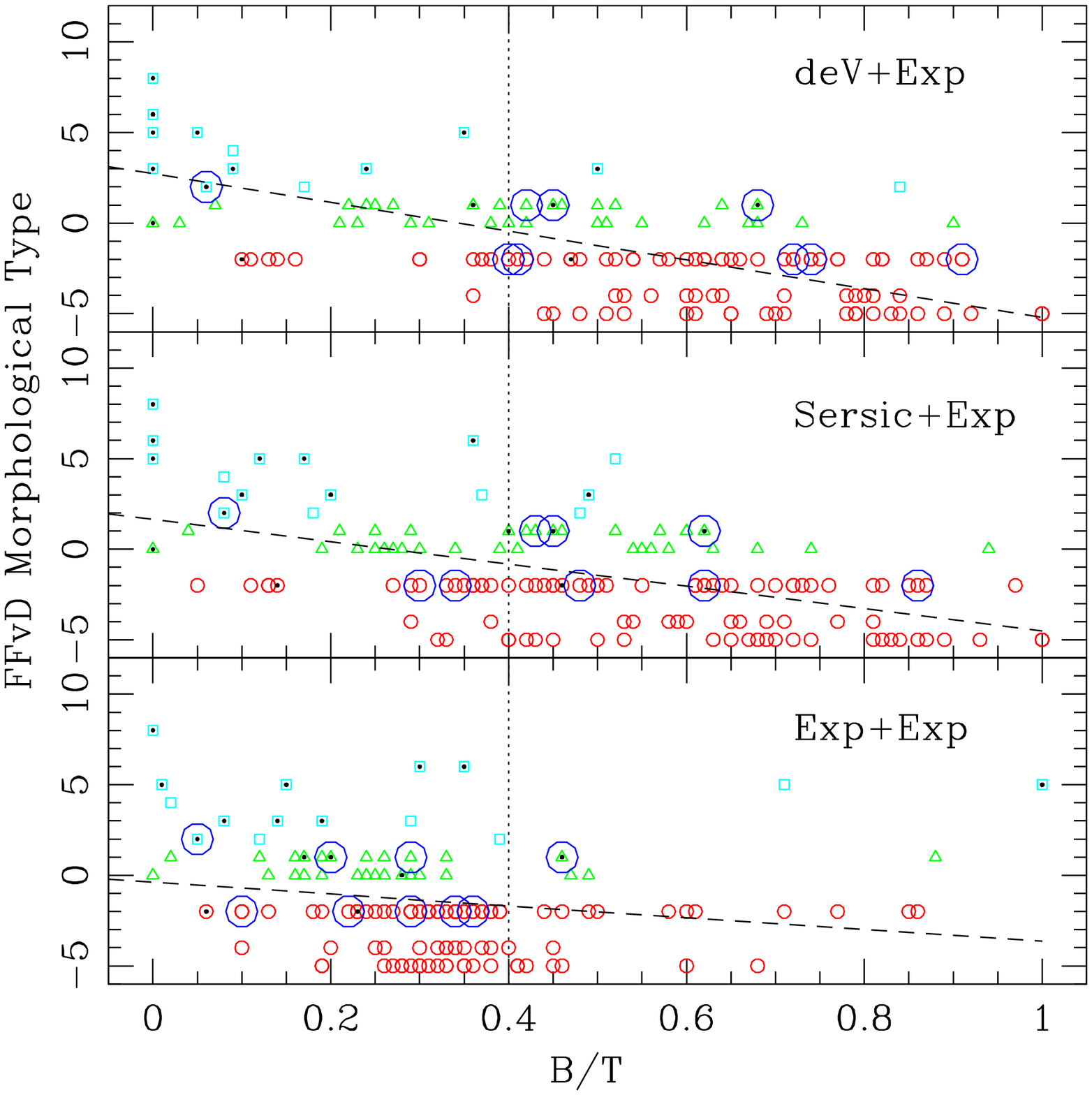}
\caption{Comparison of bulge-to-total fractions to Hubble types from
  FFvD00 for 155 typed cluster members ($m_{814}\leq21$,
  $MB_z\sim-17.3$\logh).  The top panel shows $B/T$ as defined using a
  de Vaucouleurs bulge+exponential disk, the middle a S\'ersic bulge
  ($0.2\leq n\leq 6$)+exponential disk, and the bottom a double
  exponential profile; all profiles are measured in the F814W filter
  ($\sim$rest-frame $V$).  The small open circles represent E-S0's
  ($-5\leq T\leq-1$), open triangles S0-a ($0\leq T\leq1$), open
  squares spirals and irregulars ($2\leq T\leq15$), and large open
  circles post-starburst galaxies; members with a high degree of galaxy
  asymmetry ($R_A\geq0.05$) have a solid dot.  The vertical line
  denotes the adopted break between bulge ($B/T\geq0.4$) and disk
  ($B/T<0.4$) dominated galaxies \citep{tran:01,im:01}.  Both the de
  Vaucouleurs+exponential disk and S\'ersic bulge+exponential disk
  correspond well to visual morphologies, albeit with large scatter;
  the trend between $B/T$ and Hubble type for both profiles is 99\%
  significant ($>2\sigma$) using the Spearman rank test.
\label{1358profiles}}
\end{figure}

A fundamental question when using parametric surface brightness models
to measure physical properties is what type of profile should be used.
Depending on the profile, certain structural parameters (\eg~$B/T$,
scale lengths, and bulge/disk colors) can change drastically in value
(see Fig.~\ref{1358profiles}).  Recent work
\citep{dejong:96,courteau:96,andredakis:98} combine two $n = 1$
exponential components to fit the surface brightness profile of
late-type spirals while others \citep{caon:93,labarbera:02} use a
S\'ersic profile to fit early-type galaxies.  Another popular profile
is the classical de Vaucouleurs with exponential disk
\citep{schade:95,marleau:98,simard:99,im:01,tran:01}.  If conclusions
are to be drawn from these quantitative studies of galaxy morphology,
the most basic first step is to be consistent in the type of profile
used.

In this paper, we test three different profiles by fitting them to the
cluster galaxies and comparing the measured structural parameters to
published Hubble types from FFvD00; as we justify in \S5, we use only
the 155 members in our sample brighter than $m_{814}=21$
($MB_z\sim-17.3-5$\logh).  The three profiles are: (1) de Vaucouleurs
bulge with exponential disk; (2) S\'ersic bulge ($0.2\leq n\leq 6$)
with exponential disk; and (3) double exponential.
Figure~\ref{1358profiles} shows $B/T$ versus visual morphological type
for the three profiles; here, visual types are represented as E (-5),
S0 (-2), Sa (1), Sb (3), Sc (5), Sd (7), Sm (9), and Im (10); galaxies
with high degree of asymmetry ($R_A\geq0.05$, see \S\ref{hiRA}) are
noted with dots.

Of the three profiles, both the de Vaucouleurs bulge+exponential disk
and S\'ersic bulge+exponential disk correspond well to the visual
types.  The correlation between measured $B/T$ and visual type for both
profiles is 99\% ($>2\sigma$) significant with the Spearman rank test
\citep{press:92}.  The trend between de Vaucouleurs bulge+exponential
disk and type is not surprising because: 1) bright ellipticals and the
bulges of early-type spirals are well-fit by a de Vaucouleurs profile
\citep{andredakis:95,andredakis:98}; and 2) cluster galaxies tend to be
bulge-dominated systems.  We note, however, that there is large scatter
in the correlation and that $B/T$ cannot reliably distinguish
early-type spirals from ellipticals and S0's.

\begin{figure}
\plotone{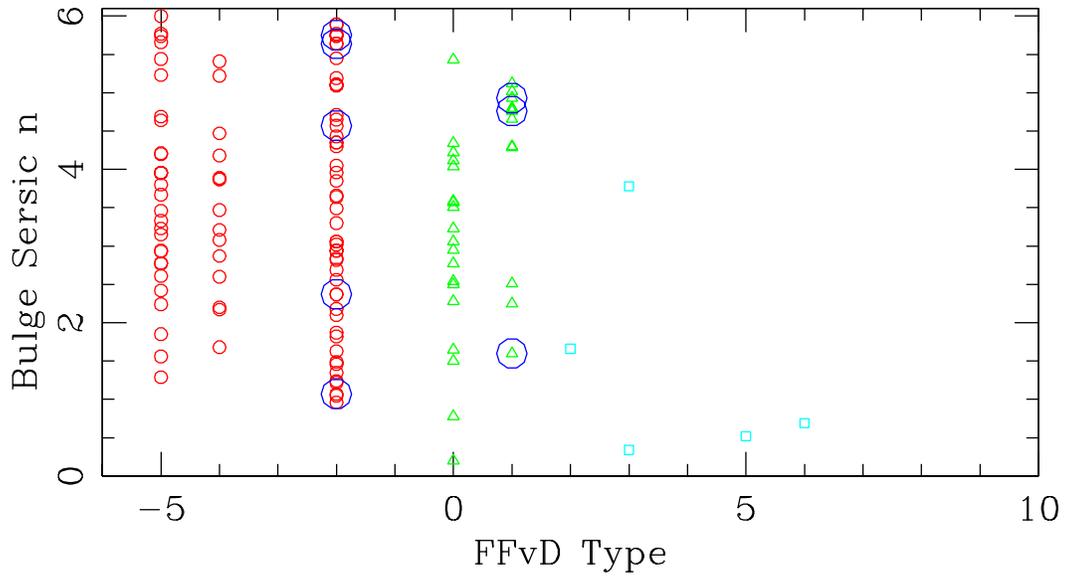}
\caption{Comparison of Hubble types from FFvD00 to the best fit bulge 
power $n$ ($0.2\leq n\leq6$) for the S\'ersic bulge+exponential disk 
model; here we consider only members brighter than our cut-off magnitude
($m_{814}\leq21$) with $(B/T)_{Ser}\geq0.25$.  In our cluster sample,
we find no correlation between S\'ersic index $n$ and visual type.
\label{1358n_type}}
\end{figure}

While the correlation between visual type and $B/T$ for the S\'ersic
bulge+exponential disk fits is strong (see Fig.~\ref{1358profiles}), we
do not recover the observed correlation between bulge power $n$ and
Hubble type found locally \citep[see
Fig.~\ref{1358n_type}]{caon:93,graham:99}.  This combined with results
from the simulations discussed in \S6 suggest that a S\'ersic
bulge+exponential disk is not an appropriate model for this sample.
Our sample does not contain enough spirals to test for the trend
between $n$ and early versus late-type spirals found by
\citet{graham:01}.

There is no visible correlation between $B/T$ and type for the double
exponential profile.  The mismatch between profile and galaxy type
results in an alarming number of cluster galaxies (75\%) being
classified as disk-dominated sytems ($B/T<0.4$).  The lack of
correlation is not surprising, however, given the inherent ambiguity
between ``bulge'' and ``disk'' when describing the two components with
the same functional form.  Also, this profile is appropriate mainly for
late-type spirals \citep{dejong:96,courteau:96}.  These points convince
us that the double exponential profile is a poor choice for measuring
structural parameters across the range of galaxy types.

Based on these tests, we proceed to use the de Vaucouleurs bulge with
exponential disk profile as our canonical fitting model for the sake
of continuity across the full range of morphological types. 

\subsection{2D Surface Brightness Models:  Sanity Checks}

Before drawing any conclusions from the surface brightness models, we
employ ``sanity checks'' to ensure our method corresponds to observed,
well-established cluster galaxy properties.  These tests are: 1) check
for consistency of $B/T$ measurement between filters; 2) establish a
correlation between measured $B/T$ and visual morphological type; 3)
determine if a morphology-radius ($\sim$ density) relation using $B/T$
exists; and 4) check that typing cluster members by $B/T$ results in
separation of galaxy types in the color-magnitude diagram.

\subsubsection{Galaxy Models:  de Vaucouleurs Bulge with Exponential
Disk} 

\begin{figure}
\epsscale{0.8}
\plotone{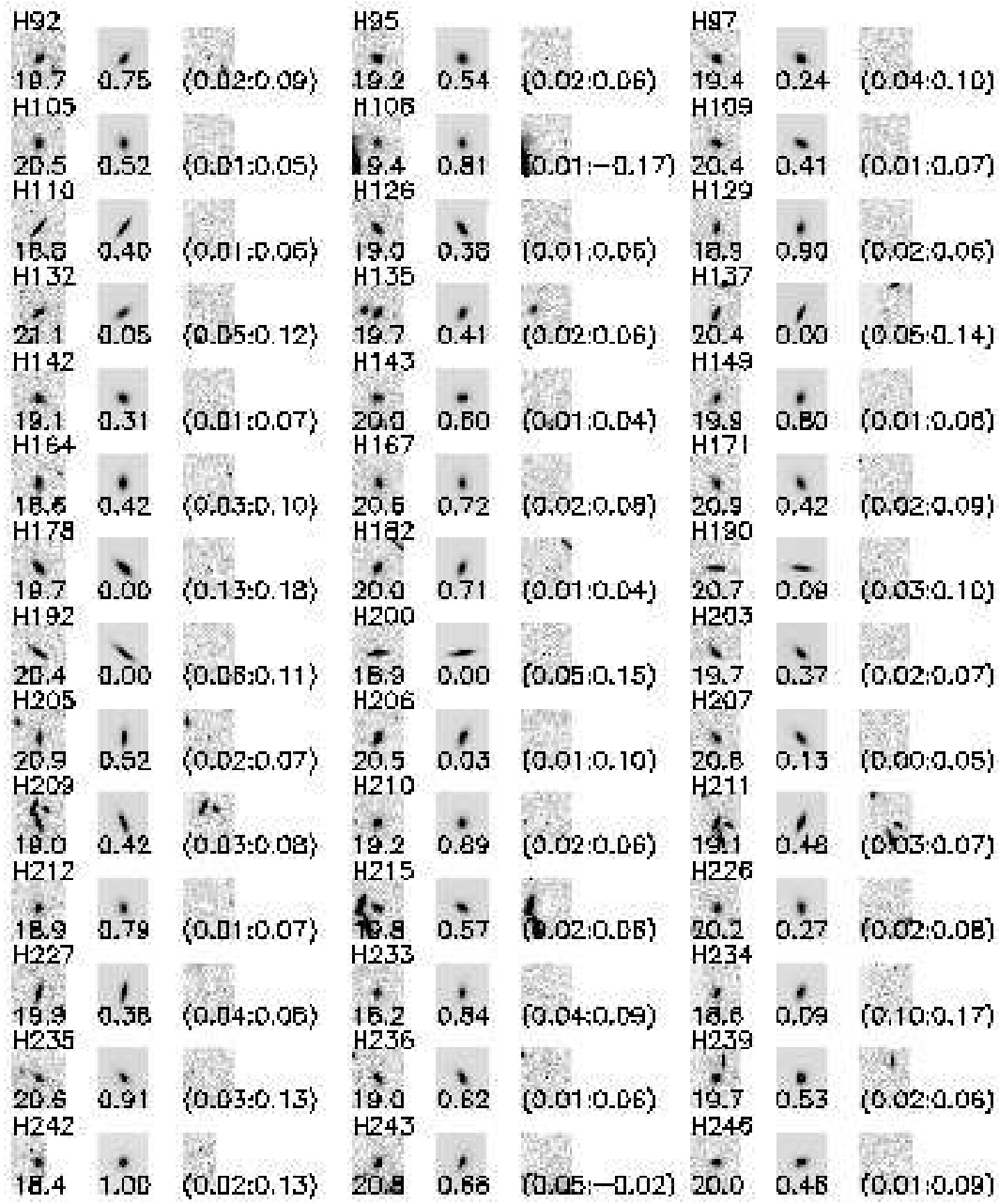}
\caption{Thumbnail images of 168 cluster members in our sample of 173 fitted
  galaxies.  These images are from the F814W filter ($\sim V_0$).  The
  pixel area in each thumbnail is 15 times the galaxy's isophotal area
  as defined by SExtractor (typically $10-15''$ on a side).  Each set
  of three images shows the galaxy, its best-fit de Vaucouleurs bulge
  with exponential disk model, and its residual image created by
  subtracting the model from the original.  Identification numbers are
  in the upper left corner where the BCG is H375.  Included in the
  thumbnails are the apparent F814W magnitude (left), bulge/total
  luminosity ($B/T$; middle), and the asymmetry parameter and total
  fraction of residual light ($R_A, R_T$; right). {\bf Higher
  resolution images of the 168 members are available at
  http://www.exp-astro.phys.ethz.ch/vy/astronomy}
\label{1358Igals}}
\end{figure}

We fit a de Vaucouleurs bulge with exponential disk model to the 173
cluster members in our sample.  The galaxies are fit in each filter
separately to test the robustness of the structural parameter
measurements.  Figure~\ref{1358Igals} shows 168 of the 173 galaxies in
our cluster sample.  Each set of three thumbnails shows the galaxy, its
best-fit de Vaucouleurs with exponential disk model, and its residual
image created by subtracting the model from the original.  Included in
the images are reference numbers (upper left), apparent F814W
magnitudes (bottom left), bulge/total luminosity ($B/T$; bottom
middle), and the asymmetry parameter and total fraction of residual
light ($R_A, R_T$; bottom right).  The pixel area in each thumbnail is
15 times the galaxy's isophotal area as defined by SExtractor; this is
usually $10-15''$ on a side.  The cluster members range in absolute
$B_z$ magnitude from $-20.8$\logh~(BCG; H375) to $-15.9$\logh~(H1829).
Table~\ref{params} lists these parameters as well as $B_z$, $(B-V)_z$,
Hubble type, half-light radius, and bulge/disk scale lengths.  In the
following discussion, we only include the 155 members with
$m_{814}\leq21$ as, based on results from extensive simulations in \S5,
our measured structural parameters are not significantly biased for
these.

For simplicity, we focus on structural parameters measured from the
individual F814W ($\sim$rest-frame $V$) images as opposed to including
values from the F606W ($\sim$ rest-frame $B$) images.  Comparing
structural parameters measured in the two filters finds little
difference in their general distributions.  We emphasize that while
structural parameter values for individual galaxies may vary between
the two filters, these differences are 1) expected and understandable
and 2) do not impact the global conclusions drawn from our analysis.

\subsubsection{$B/T$ vs. Visual Morphology}

\begin{figure}
\plotone{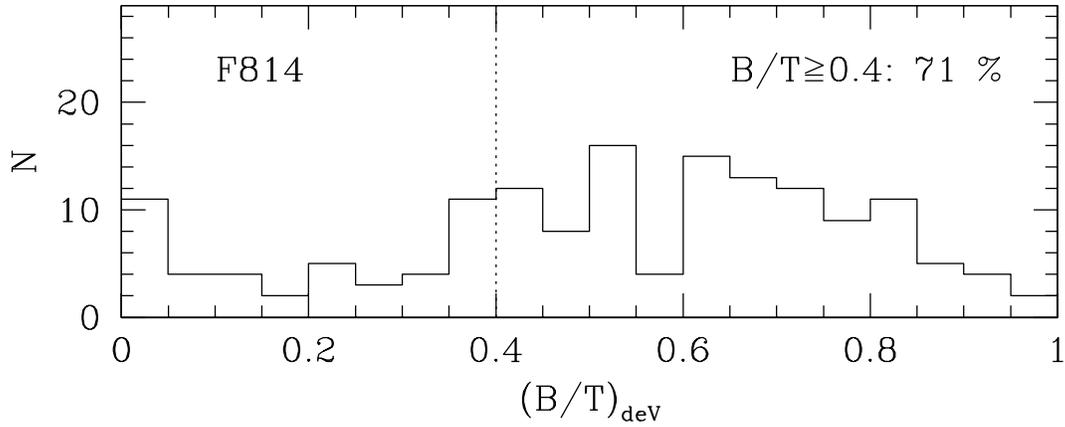}
\caption{Distribution of bulge/total luminosities [$(B/T)_{deV}$] 
measured with GIM2D in the F814W filter ($\sim V_0$) for 155 cluster members
($m_{814}\leq21$~mags).  The dotted line denotes the adopted break
between bulge ($B/T\geq0.4$) and disk ($B/T<0.4$) dominated galaxies.
We also determine the $(B/T)_{deV}$ distribution for the same galaxies 
in the F606W filter ($\sim B_0$) and find, using the K-S test, that the
distribution is not dependent on filter.
\label{1358bt_hist}}
\end{figure}

In Fig.~\ref{1358bt_hist}, we show the bulge distributions measured in
F814W.  Not surprisingly, the majority of cluster members ($\sim70$\%)
are bulge-dominated ($B/T\geq0.4$) systems.  This result does not
depend strongly on filter as, using the Kolmolgorov-Smirnov test
\citep{press:92}, we find the F814W and F606W distributions to be
indistinguishable.

With the de Vaucouleurs bulge+exponential disk profile, we find a
strong trend (99\% confidence) between $B/T$ and visual morphological
type (Fig.~\ref{1358profiles}, top panel) with the Spearman rank test
\citep{press:92}, albeit with large scatter.  We also see in
Fig.~\ref{1358profiles} that of the high asymmetry, visually typed
galaxies, all have $B/T<0.7$; this is expected because disks are more
likely to harbor asymmetric features such as HII regions.  The
correlations between $B/T$, asymmetry, and visual type are highly
encouraging but we note that, in this sample, it is difficult to
separate early-type spirals from ellipticals and S0's using $B/T$
alone.

\subsubsection{$B/T$ (Morphology) vs. Local Density}

\begin{figure}
\plotone{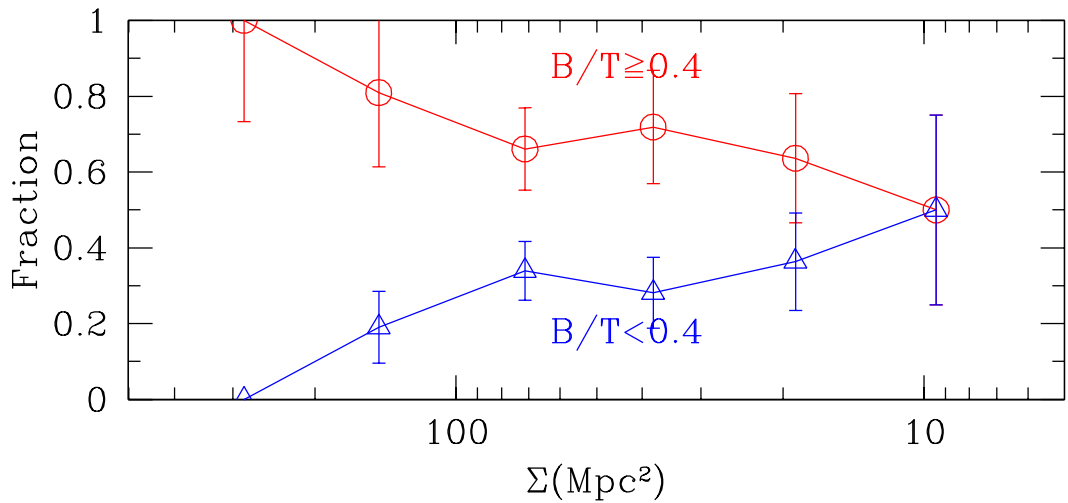}
\caption{Number fraction of bulge [$(B/T)_{deV}\geq0.4$; open circles] and
disk-dominated [$(B/T)_{deV}<0.4$; open triangles] galaxies as a function of
local density for the 155 cluster members ($m_{814}\leq21$~mags);
Poisson errorbars are shown.  As is observed in lower clusters
\citep{dressler:80}, the fraction of disk-dominated galaxies increases
with decreasing local density.
\label{1358frac_R}}
\end{figure}

F98 and vD98 have shown that galaxy spectral type and visual
morphological type vary as a function of local density in CL~1358+62.
In Fig.~\ref{1358frac_R}, we show that fitting a de Vaucouleurs bulge
with exponential disk profile to the same sample of galaxies also
recovers the same results, as expected since we have demonstrated the
strong correlation between $B/T$ and visual type.  Splitting the
sample into bulge and disk-dominated systems, Fig.~\ref{1358frac_R}
shows the increase in the fraction of disk-dominated systems with
decreasing local galaxy density.  Here, the local galaxy density is
defined as $\Sigma=11/\pi r_{10}^2$, and $r_{10}$ is the distance to
the farthest of the ten nearest confirmed members to the object
\citep{dressler:80}.

\begin{figure}
\plotone{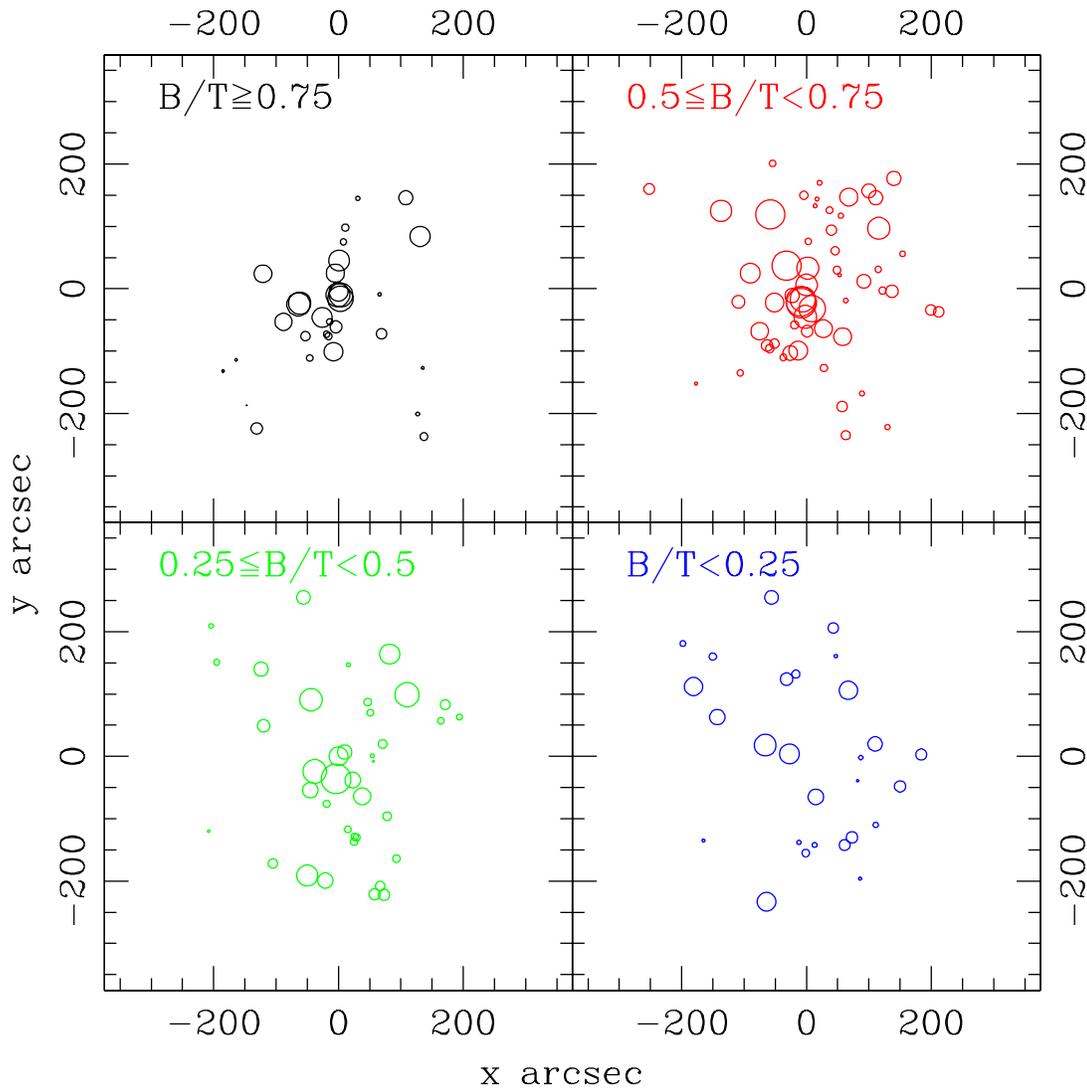}
\caption{Clustering, as defined by the Dressler-Schectman test
\citep{dressler:88}, as a function of $(B/T)_{deV}$ for cluster members
($m_{814}\leq21$~mags).  The scale is $50''\sim225$\hi kpc. The
circles are proportional to how much the local mean redshift and
velocity dispersion deviate from that of the cluster's; many large
circles grouped together indicate significant substructure.  As we
move from bulge to disk-dominated systems (clockwise from upper left
panel), we find that bulge-dominated systems are clustered more
strongly than disk-dominated ones.
\label{1358ds_bt}}
\end{figure}

An interesting question is whether the distribution of bulge-dominated
galaxies is related to cluster substructure.  Using the
Dressler-Shectman test \citep{dressler:88}, F98 found the degree of
substructure to be significant with 96\% confidence ($>2\sigma$) where
the significance was calibrated with $10^3$ Monte Carlo realizations.
Fig.~\ref{1358ds_bt} shows the spatial distribution of
$m_{814}\leq21$~mags members as a function of $B/T$ and $\delta$, the
substructure statistic.  As $\delta$ quantifies how much the local mean
redshift and velocity dispersion deviate from the cluster's global
values, groups of large circles indicate a significant degree of
substructure.  Moving from bulge to disk-dominated systems, we find
that bulge-dominated systems are clustered more strongly than
disk-dominated ones.  The spatial distribution of the most
bulge-dominated galaxies ($B/T\geq0.75$) is different from that of the
most disk-dominated ($B/T<0.25$) ones with 97\% confidence ($>2\sigma$)
using the two dimensional K-S test.  The most significant subclump is a
foreground group composed mainly of early-types ($B/T\geq0.4$),
including the BCG.

As a further refinement, we separate the cluster sample into
$B/T\geq0.5$ and $B/T<0.5$ members and apply the Dressler-Schectman
test; again, the substructure signficance is calibrated with $10^3$
Monte Carlo realizations.  The bulge-dominated galaxies (98) show
substructure with 98\% confidence ($>2\sigma$) while the disk-dominated
galaxies (75) lack significant substructure.  In addition, the cluster
velocity dispersion determined from bulge-dominated systems is lower
than that from disk-dominated ones: $950\pm70$ \kms compared to
$1220\pm100$ \kms.

\subsubsection{Color-Magnitude Relation}

\begin{figure}
\plotone{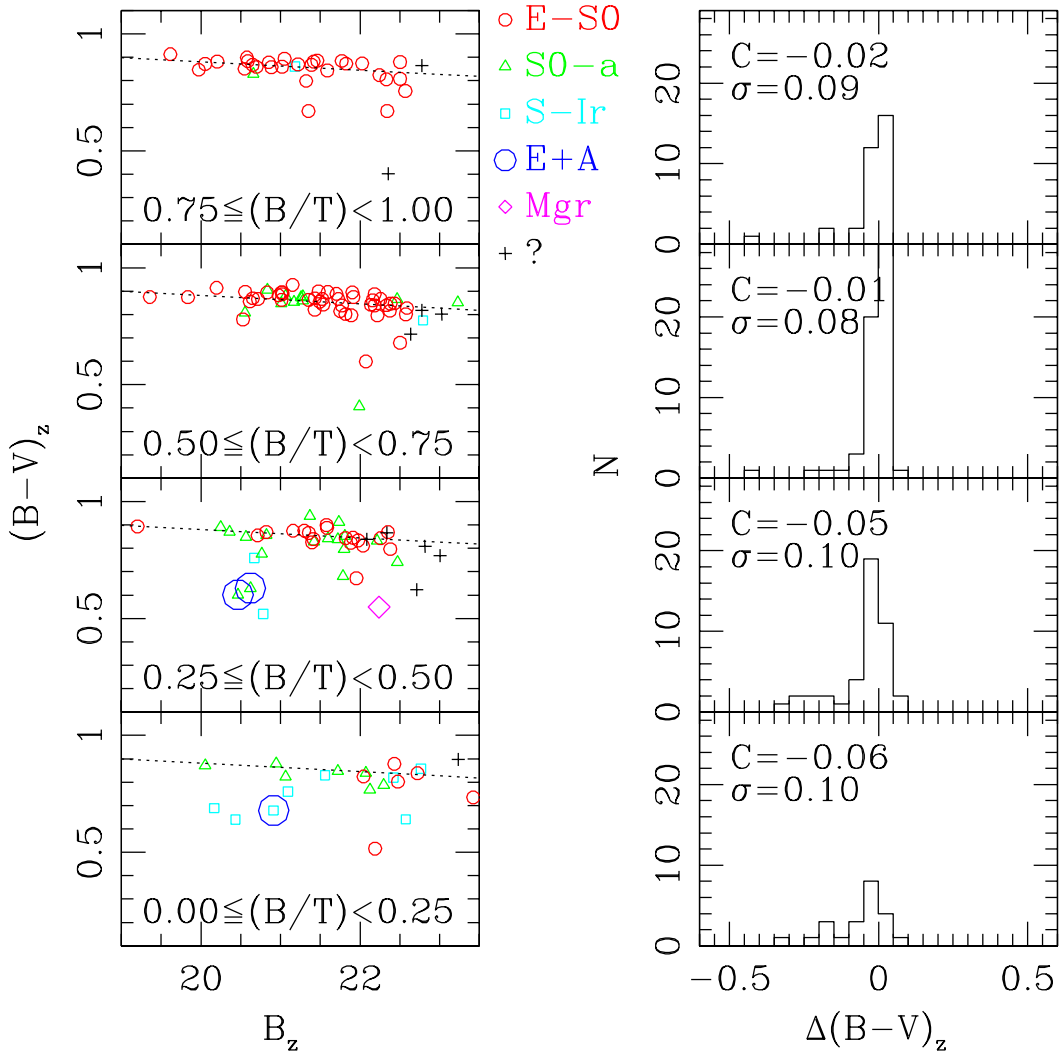}
\caption{Color-magnitude diagram for different galaxy types as defined by
$(B/T)_{deV}$.  Open circles are E-S0's ($-5\leq T \leq-1$), open triangles
S0-a's ($0\leq T\leq1$), open squares spirals and irregulars ($2\leq
T\leq15$), plus symbols non-typed ($m_{814}>22$) members, mergers open
diamonds ($T=99$), and large open circles E+A's.  The left panels show
$(B-V)_z$ vs. $B_z$ (from vD98) for four $B/T$ bins with the most
bulge-dominated sample ($B/T\geq0.75$) at the top.  The CM relation,
as determined by vD98 from the E-S0 members, is shown as a dotted
line in each of these panels.  The right panels show the corresponding
scatter about the CM relation for the four $B/T$ bins.  Here $C$ is
the average deviation from the CM relation and $\sigma$ is the width
of the deviation distribution.
\label{1358cmd}}
\end{figure}

Following vD98, we examine how galaxy colors change as a function of
magnitude and galaxy type (as defined by bulge fraction $B/T$).  In
Fig.~\ref{1358cmd}, we show the color magnitude diagram in four
$B/T$ bins; we use $B_z$ magnitudes, $(B-V)$ colors, CM relation
(determined from E-S0 members) from vD98.  As $B/T$ decreases, the
scatter and average deviation from the CM relation increases.  Also,
the most disk-dominated systems ($B/T<0.25$) are significantly bluer
and have the largest scatter about the CM relation.  These results are
consistent with the trends vD98 found using Hubble classifications
for the same galaxies.  

In Fig.~\ref{1358cmd}, we see there exist cluster members with
significant disk components ($B/T<0.5$) that are also red.  It is
possible these disk systems have had their star formation
``strangled'' \cite{balogh:98} and/or they have managed to avoid major
galaxy interactions that would have disrupted their disks,
\eg~mergers.  An extensive analysis of the bulge/disk colors of these
galaxies will be discussed in a future paper.

\subsection{Image Residuals of Cluster Members}

\subsubsection{High Asymmetry Galaxies}\label{hiRA}

Following \citet{tran:01} and \citet{schade:95}, we define galaxies
with $R_A\geq0.05$ as having a high degree of asymmetry.  In the
cluster sample of galaxies with $m_{814}\leq21$, $\sim10$\% have
$R_A\geq0.05$, and all of these high asymmetry galaxies have $B/T<0.7$.
Most asymmetries in this population are due to star-forming regions in
disk-dominated systems (\eg~H178, H200, H234, and H396; see
Fig.~\ref{1358rart}).

CL~1358+62's fraction of high asymmetry galaxies is more similar to
that observed in groups than in the field.  \citet{tran:01} found that
about 11\% of galaxies nearby, X-ray luminous groups have high
asymmetry\footnote{Note, however, that our $HST$ data are $\sim3\times$
  higher in physical resolution than the ground-based sample (0.44\hi
  kpc versus 1.31\hi kpc).}, with the fraction of bulge-dominated
galaxies in groups at $\sim50-60\%$.  In comparison, \citet{schade:95}
found that 10/32 (30\%) $HST$ observed field galaxies within the
redshift range $0.5<z<1.2$ have high galaxy asymmetry.  This is as
expected, however, as the majority of field galaxies are
disk-dominated, and disks are more sensitive to morphological
disturbances.

\subsubsection{High Total Residual Galaxies}

\begin{figure}
\epsscale{0.5}
\plotone{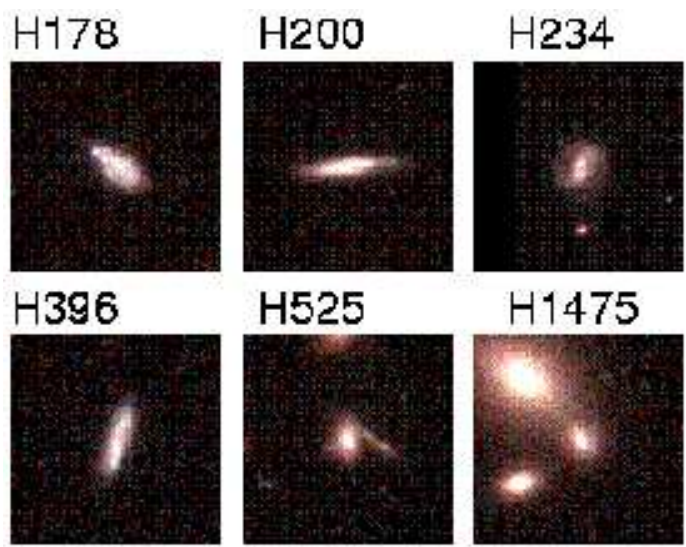}
\caption{Examples of CL~1358+62 members with a high degree of galaxy
  asymmetry ($R_A\geq0.05$) and/or high total residual ($R_T\geq0.1$).
  H525 is a galaxy with low $R_A$ and high $R_T$, while H1475 has high
  $R_A$ and low $R_T$.  The other four galaxies have both $R_A\geq0.05$
  and $R_T\geq0.1$.  {\bf Higher resolution image is available at
  http://www.exp-astro.phys.ethz.ch/vy/astronomy/}
\label{1358rart}}
\end{figure}

After examining the residual images (Fig.~\ref{1358Igals}), we define
galaxies with $R_T\geq0.1$ as having high total residual.
Approximately 20\% of the members have high total residuals with
features that are over or under-subtracted by the model, \eg~bars
(H525, see Fig.~\ref{1358rart}) or star-forming regions (H178, see
Fig.~\ref{1358rart}).  As there are 15 galaxies common to both
populations, however, this is not surprising.

We find that disk-dominated galaxies tend to be more disturbed than
bulge-dominated ones: a third of these systems show high asymmetry and
half have a high total residual.  Comparatively, the early-type
galaxies have smoother galaxy profiles: only $\sim5$\% show high galaxy
asymmetry and $\sim10$\% have a high total residual.  Normalizing the
residuals by the average surface brightness instead of the total flux
finds the same result.  The difference in the high residual fraction of
disk-dominated members compared to bulge-dominated ones supports
findings that the mechanism(s) responsible for morphologically
disrupting cluster members are: (1) more effective at disturbing
disk-dominated galaxies than bulge-dominated ones and/or (2) the
early-type galaxies are older and have been in the cluster environment
longer, so any disruptions in these galaxies occurred long ago and have
damped out via phase mixing.

\subsection{Correlations between Structural Parameters and
[OII]$\lambda3727$ Emission}

To determine how morphological characteristics are correlated (or not)
with ongoing star formation, we compare $B/T$, $R_A$, and $R_T$ to
[OII]$\lambda3727$ measurements from F98.  We assume [OII] emission is
correlated with star formation and not active galactic nuclei (AGN) as
we do not have the necessary data (\eg~H$\alpha$ or X-ray) to
distinguish between the two.  Again, we discuss only the 155 galaxies
in the cluster sample with $m_{814}\leq21$.  In the following
discussion, note that neither $B/T$ nor $R_A$ are correlated with
luminosity, as determined with the Spearman rank test.

\subsubsection{Current Star Formation vs. $B/T$}

\begin{figure}
\epsscale{0.8}
\plotone{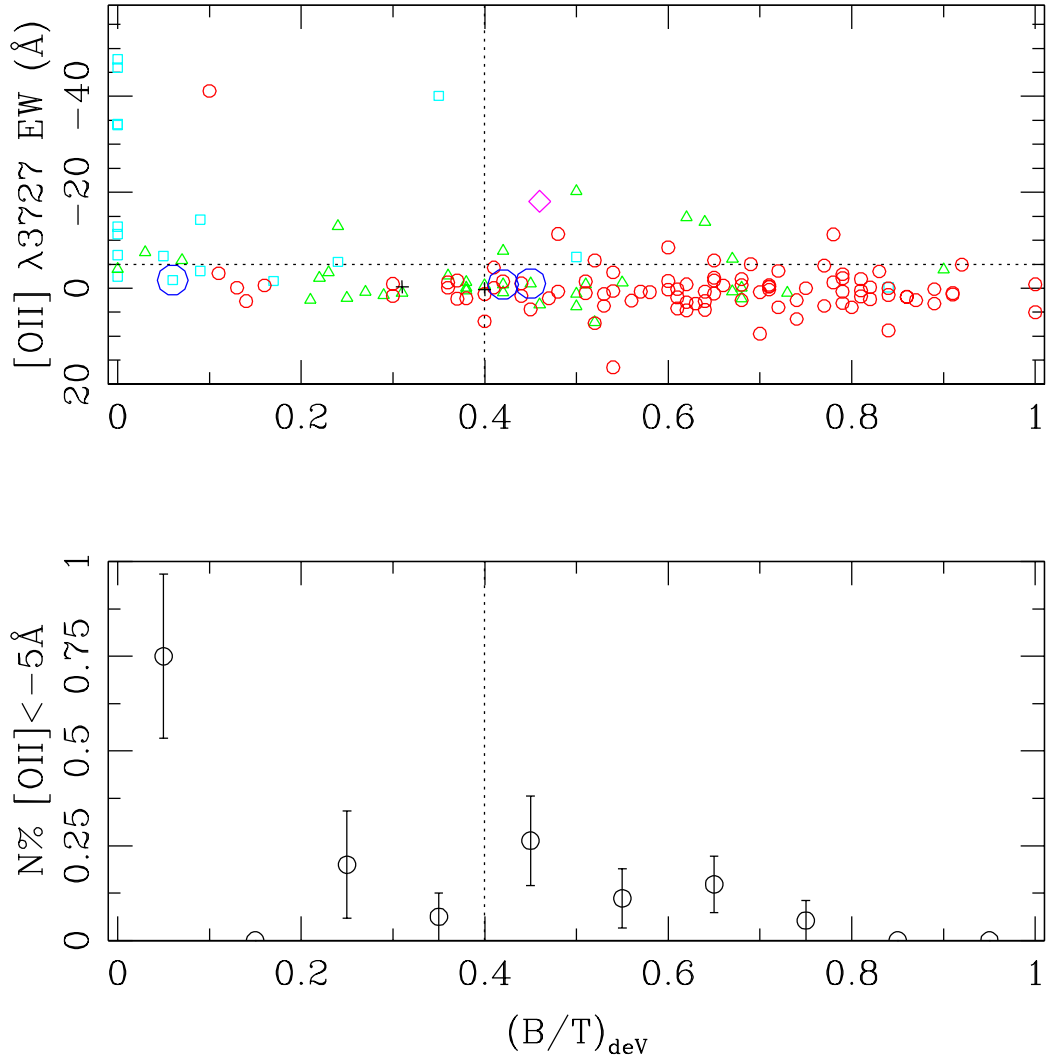}
\caption{{\it Top Panel:} Distribution of $(B/T)_{deV}$
vs. [OII]$\lambda3727$ equivalent width (F98) of cluster members
($m_{814}\leq21$); negative [OII] EW values correspond to emission.
The vertical and horizontal dotted lines refer to $B/T=0.4$ and [OII]
EW$=-5$\AA, the adopted division between bulge/disk-dominated and
star-forming ([OII]$<-5$\AA) respectively.  Symbols are as in
Fig.~\ref{1358cmd}.  {\it Bottom Panel:} The fraction of galaxies in
each $B/T$ bin with [OII] EW$<-5$ where the dotted line denotes
$B/T=0.4$.  Pure disk systems have a high probability of having
significant [OII]$\lambda3727$ emission.
\label{1358OII_bt}}
\end{figure}

Figure~\ref{1358OII_bt} shows the distribution of the galaxies' bulge
fractions versus their [OII] equivalent widths; negative [OII] EW
values correspond to emission.  Of the 26/155 galaxies with
significant [OII] emission ([OII] EW$<-5$\AA), $\sim60$\% are
disk-dominated galaxies ($B/T<0.4$).  Even more compelling, 4 of the 7
galaxies with the highest star formation rates ([OII]$\leq-20$\AA)
have $B/T=0$, $i.e.$, completely disk-dominated systems.

Current star formation in cluster galaxies, as traced by
[OII]$\lambda3727$ emission, is not confined to only disk-dominated
systems (see bottom panel, Fig.~\ref{1358OII_bt}).  Interestingly,
$\sim40$\% of galaxies with significant [OII] emission are
bulge-dominated ($B/T\geq0.4$).  These galaxies span the range of
luminosity ($\sim$ mass), having $-17.3\geq MB_z-5\log
h\geq-20.8$, and include the BCG.  We discuss this population in
more detail in \S\ref{1358agn}.

Comparison of the [OII] emission line strength to bulge fraction finds
it to be correlated strongly with $B/T$.  Table~\ref{OIItab} shows the
average [OII] EW in four $B/T$ bins.  The average [OII] emission is
dramatically higher for galaxies with $B/T<0.25$, dropping to zero for
$B/T\geq0.85$ galaxies.  Like \citet{balogh:98}, we find that the most
bulge-dominated members ($B/T\geq0.8$) have [OII] EQW consistent with
zero.

\subsubsection{Current Star Formation vs. Galaxy Residuals}

\begin{figure}
\plotone{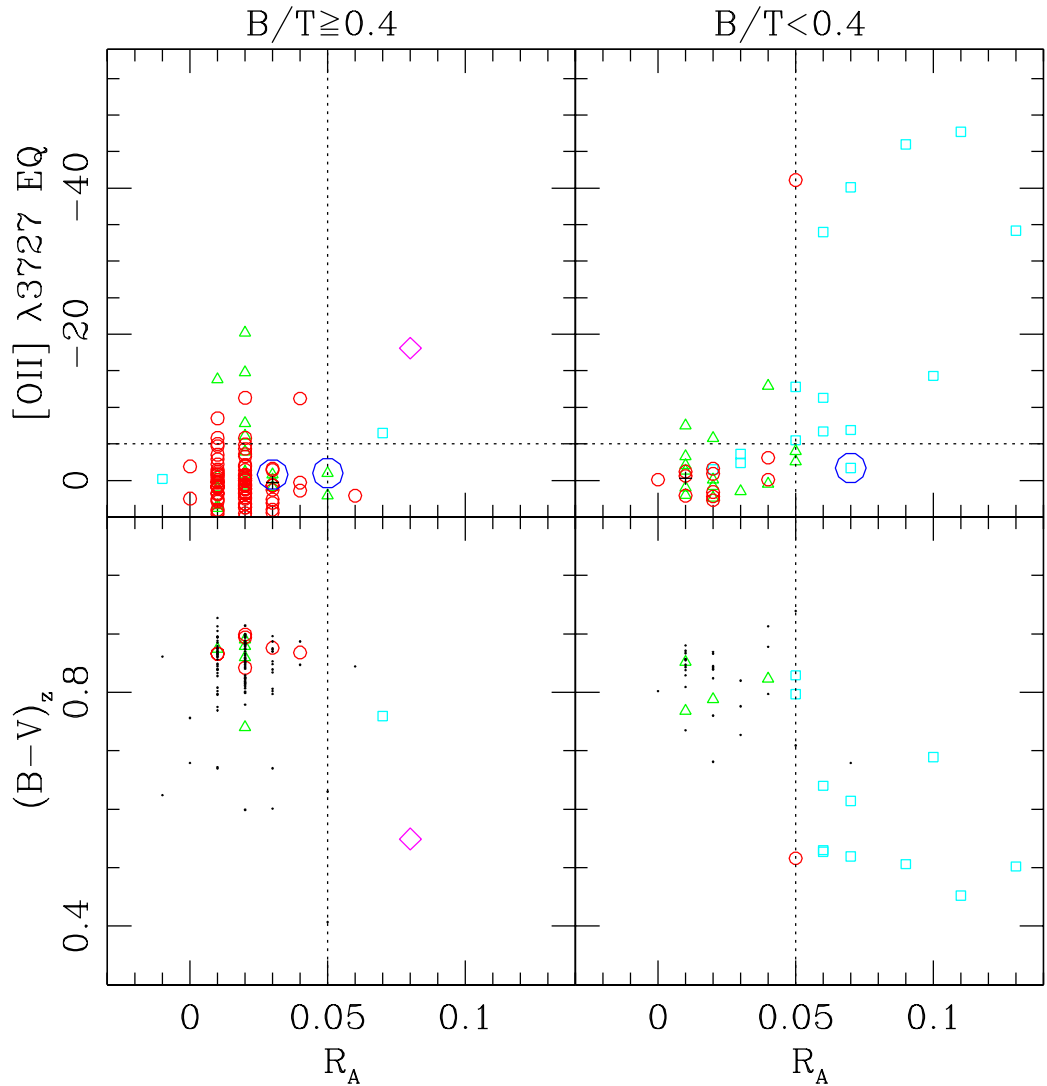}
\caption{{\it Top Panels:} Correlation between high galaxy asymmetry
($R_A\geq0.05$) and strong [OII] emission ($<-5$\AA) for
bulge-dominated (top left) and disk-dominated (top right) members; only
galaxies with $m_{814}\leq21$ mags are included, and symbols are as in
Fig.~\ref{1358cmd}.  Vertical and horizontal dotted lines show the
adopted cuts for high asymmetry and star-forming ([OII]$<-5$\AA)
galaxies respectively.  The trend of increasing [OII]$\lambda3727$
emission with $R_A$ for the {\it disk-dominated} galaxies has 99\%
confidence with the Spearman rank test.  There is no detectable trend
for the bulge-dominated galaxies. {\it Bottom Panels:} Distribution of
colors for the two populations where members with no significant [OII]
emission are shown as dots, and members with [OII]$<-5$\AA~are shown as
symbols.  Virtually all of the S0-a members with significant [OII] are
red.
\label{1358OII_RA}}
\end{figure}

The correlation between high galaxy asymmetry and strong
[OII]$\lambda3727$ emission for bulge and disk-dominated galaxies is
shown in Fig.~\ref{1358OII_RA}.  Approximately 70\% of the high
asymmetry galaxies have significant [OII] emission.  For the
disk-dominated members, there is a strong trend (99\% confidence) of
increasing [OII] emission with increasing $R_A$; most of these
galaxies have high asymmetry and/or high total residuals and are blue.
We find that current star formation in the disk-dominated population
is highly correlated with measureable morphological disturbances.

Although 11 of the bulge-dominated galaxies do have strong [OII]
emission, only three of these galaxies have $R_A\geq0.05$ and/or
$R_T\geq0.1$.  Assuming the [OII] emission is due to star formation,
active star formation tends to be distributed in a smooth and uniform
manner when it occurs in $B/T\geq0.4$ members.  Alternatively, these
members may harbor active galactic nuclei \citep{martini:02}; we
discuss this possibility in greater detail in \S\ref{1358agn}.  Using
the high total residual population instead of the high asymmetry one,
we find essentially the same results.


\section{Towards Understanding Different Galaxy Populations}

\subsection{E's vs. S0's:  A Matter of Viewing Angle?}

\citet{jorgensen:94} suggest that $M_B>-22$ ellipticals and S0's form
the same galaxy class, such that ellipticals tend to be face-on
members of this class while S0's are more edge-on members.  Our
measurements of bulge ellipticity and disk inclination enable us to
test this hypothesis.  If the ellipticity and inclination
distributions of these two populations are completely different but
they span the same range in other physical properties, this would be
strong evidence for these ellipticals being face-on S0's.  However, as
the most luminous E/S0 galaxies (\eg~BCGs) are hypothesized to have
different formation histories \citep{kormendy:96,naab:99}, we restrict
our analysis to $19\leq m_{814}\leq21$.

\begin{figure}
\plotone{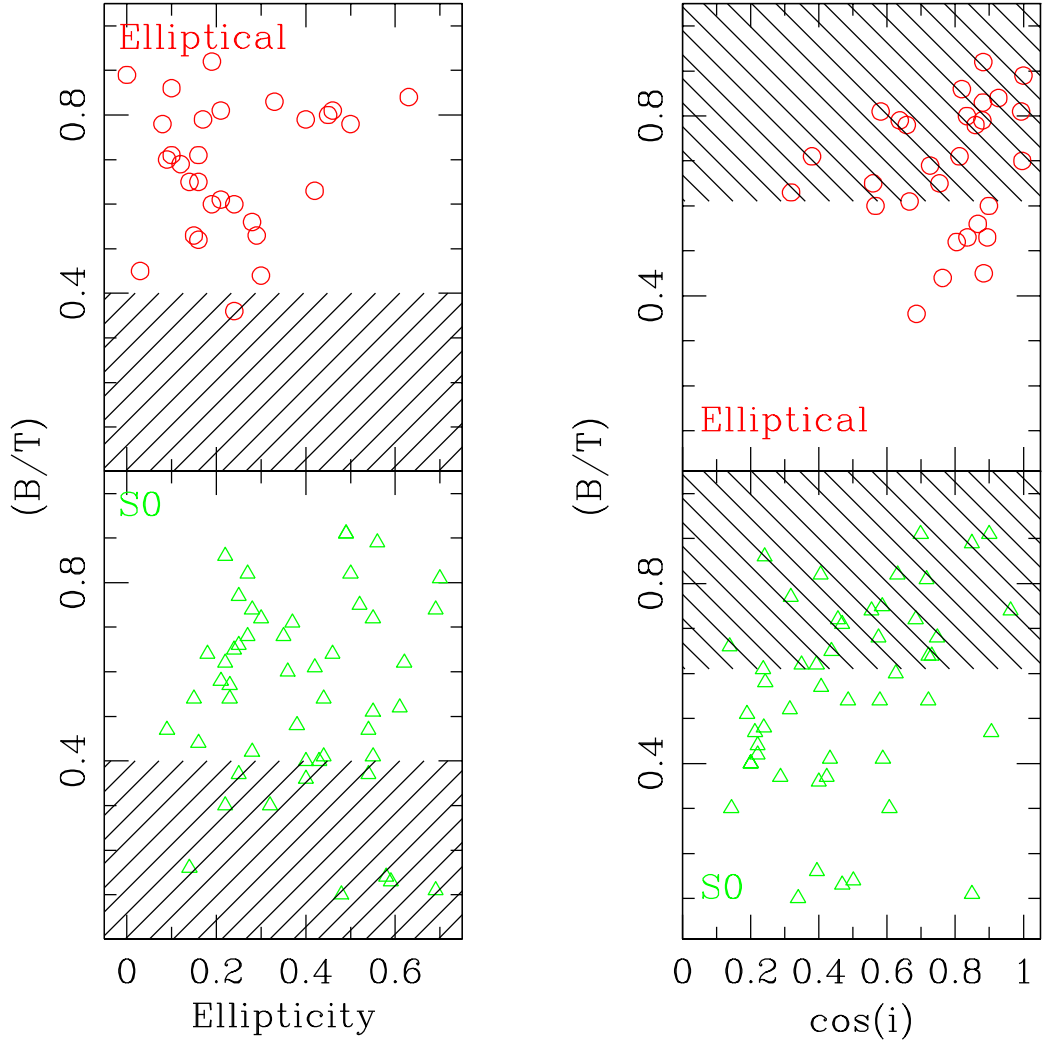}
\caption{{\it Left:} The bulge ellipticity of visually typed
ellipticals (top) and S0's (bottom; FFvD00) versus measured
$(B/T)_{deV}$; here we apply an upper magnitude cut so that $19\leq
m_{814}\leq21$.  We exclude galaxies in the hatched regions
[$(B/T)_{deV}\leq0.4$] as the random error associated with ellipticity
measurements for these objects is $\delta (Ell)\geq0.1$.  {\it Right:}
Disk inclination versus $(B/T)_{deV}$ where only $B/T\leq0.6$ members
are included in the analysis.  If E's and S0's truly are two different
populations, they should both span the same range in ellipticity and
inclination.  They do not (99\% confidence with the K-S test), and
there is a conspicuous lack of round ($Ell<0.2$) S0's.  These points
strongly suggest these ellipticals are likely to be face-on S0's.
\label{1358bt_ell}}
\end{figure}

Classically, ellipticals can be circular (E0; axis ratio is 1) to
cigar-shaped (E6; axis ratio is 0.4).  Thus, visually typed
ellipticals in CL~1358+62 (FFvD00) should span the range in bulge
ellipticity and, for systems with disks, inclination; the same is true
for S0 and even Sa galaxies.  To test this, we plot in
Fig.~\ref{1358bt_ell} the bulge ellipticity and disk inclination
versus measured $B/T$ of E's and S0's.  We include all visually typed
E's and S0's in Fig.~\ref{1358bt_ell} for completeness but note that
ellipticity and inclination measurements are robust only for
$B/T\geq0.4$ and $B/T\leq0.6$ systems respectively.

Applying a K-S test to both the bulge ellipticity and disk inclination
distributions of E's and S0's confirms they are different, yet their
$B/T$ distributions (for $B/T\geq0.4$) are indistinguishable.  Note
also the conspicuous lack of round ($Ell<0.2$) S0's, a population that
should exist due to projection effects alone.  To strengthen our
argument, we compare the internal velocity dispersion\footnote{From
\citet{kelson:00b} we have $\sigma_{disp}$ for four ellipticals and 11
S0's; for the rest, we use derived velocity dispersions
\citep{tran:03b}.}, half-light radius, projected cluster radius,
[OII]$\lambda$3727, magnitude, and $(B-V)_z$ distributions of the E's
and S0's.  Using the K-S test, we find the distributions to be
indistinguishable for all these parameters.

The fact that visually typed E's and S0's share similar properties
with the {\it exception} of bulge ellipticity and disk inclination
strongly suggests that these ellipticals are likely to be face-on
S0's.  In addition, the range in $B/T$ for members classified as
ellipticals (Fig.~\ref{1358bt_ell}, top panel) shows that these are
not pure $r^{1/4}$ galaxies.  This supports results from recent
observations and modeling \citep{rix:99,burkert:01} that a disk
component exists in most ellipticals.

\subsection{AGN's in Cluster Ellipticals?}\label{1358agn}

We find $\sim5$\% of the total cluster population are bulge-dominated
galaxies ($B/T\geq0.4$) with significant [OII]$\lambda3727$ emission
($<-5$\AA) that, unlike their disk-dominated counterparts, display
little morphological disruption ($R_A<0.05$).  A possible explanation
for these unusual systems is that they are similar to the blue field
ellipticals studied by \citet{menanteau:99,menanteau:01}, $i.e$
bulge-dominated systems still forming stars in their cores.  In this
cluster, however, the majority of these systems are {\it as red} as the
absorption population (see Fig.~\ref{1358OII_RA}) even though their
average [OII]$\lambda3727$ emission is significantly higher
($-8.5\pm3.9$~\AA~vs. $0.70\pm1.6$\AA).  Their bulge components are
equally as red, $i.e.$ these objects do not have blue cores and it is
highly unlikely that their [OII] emission is due to a central
starburst. 

Our sample finds that the undisturbed, high [OII] emission,
bulge-dominated members are not low mass objects.  Comparison of
internal velocity dispersions measured for half of this population
(6/11) from \citet{kelson:00b} shows $\sigma_{disp}$ ranges from
$128\pm5$ km/s to $230\pm5$ with the BCG at $307\pm7$ km/s.  While it
could be argued that velocity dispersions are measured most easily for
massive galaxies, thus introducing a bias exists against low mass
($\sim\sigma_{disp}$) objects, the sample published by
\citet{kelson:00b} includes robust velocity dispersions to
$\sigma\sim60$km/s.  

These points suggest that the [OII] emission in these red,
bulge-dominated galaxies originates from AGN activity rather than star
formation.  In Abell 2104, \citet{martini:02} find a surprisingly high
number of X-ray sources associated with red cluster galaxies; their
lower limit on the AGN fraction of 5\% is remarkably similar to our
cluster fraction of low asymmetry, [OII] emitting, bulge-dominated
members.  Unlike the AGN candidates in this cluster, however, the
majority of \citet{martini:02}'s AGN candidates lack significant
optical emission.  It may be that the number of AGN candidates
identified via [OII] emission only places a lower limit on the cluster
AGN fraction.

\subsection{E+A Galaxies}

Following F98, we select post-starburst (``E+A'') galaxies as having no
[OII]$\lambda3727$ emission ($\geq-5$\AA) and strong Balmer absorption
[(H$\beta+$H$\delta+$H$\gamma)/3\geq4$~\AA).  The nine
spectroscopically confirmed E+A galaxies on the HST mosaics span the
range in bulge fraction ($0<(B/T)\leq0.9$) but these are not bright
galaxies ($-19.6\leq MB_z-5\log h\leq-17.3$).  Virtually all of them
are {\it blue} \citep[see Fig.~\ref{1358cmd}, also][]{bartholomew:01},
and a third of them show significant asymmetry or total residual (see
Fig.~\ref{1358OII_RA}). A detailed analysis of the cluster E+A
population is presented in \citet{tran:03b}.

\section{Simulations:  Robustness of Surface Brightness Models}

To test the robustness of structural parameters determined by fitting
de Vaucouleurs bulge+exponential disk surface brightness models, we
generate several thousand artificial galaxies and apply the same
analysis.  Note that we assume throughout this section that galaxies
are described exactly by an $r^{1/n}$ bulge+exponential disk profile.
In the following, we focus on errors associated with incorrect
measurements of flux, $B/T$, bulge ellipticity, disk inclination,
half-light radius, and the scale lengths of the two profile components.
We also include short discussions on the reliability of fitting
S\'ersic and exponential bulges.

\subsection{Determining de Vaucouleurs Bulges}

Artificial de Vaucouleurs bulge+exponential disk galaxies are generated
uniformly to cover the same range in flux, bulge/total luminosity,
scale lengths ($r_d$, $r_e$), bulge eccentricity, and disk inclination
as the cluster sample in the F814W filter (3600s integration).  For
simplicity, the bulge and disk position angles are fixed at
$45^{\circ}$.  Once the smooth galaxy image is created, it is convolved
with the appropriate TinyTim PSF, Poisson noise is added, and it is
imbedded in a $30''\times30''$ WF2 image section that is free of any
detectable objects.  At this point, the artificial galaxy catalog is
analyzed in exactly the same manner as the observed sample.

By comparing measured to input values of the structural parameters for
the artificial galaxies, we can map systematic and random errors due
to the galaxy's position in the multi-dimensional structural space.
Systematic error is the median difference between the input and
measured value for a parameter, and random error is the $1\sigma$
width of the distribution associated with the median difference.
Statistics drawn from the simulations are robust across the range of
$m_{814}$ and $(B/T)_{814}$ as the artificial catalog is well-sampled
in each of these bins.

\subsubsection{Total Magnitude}

\begin{figure}
\plotone{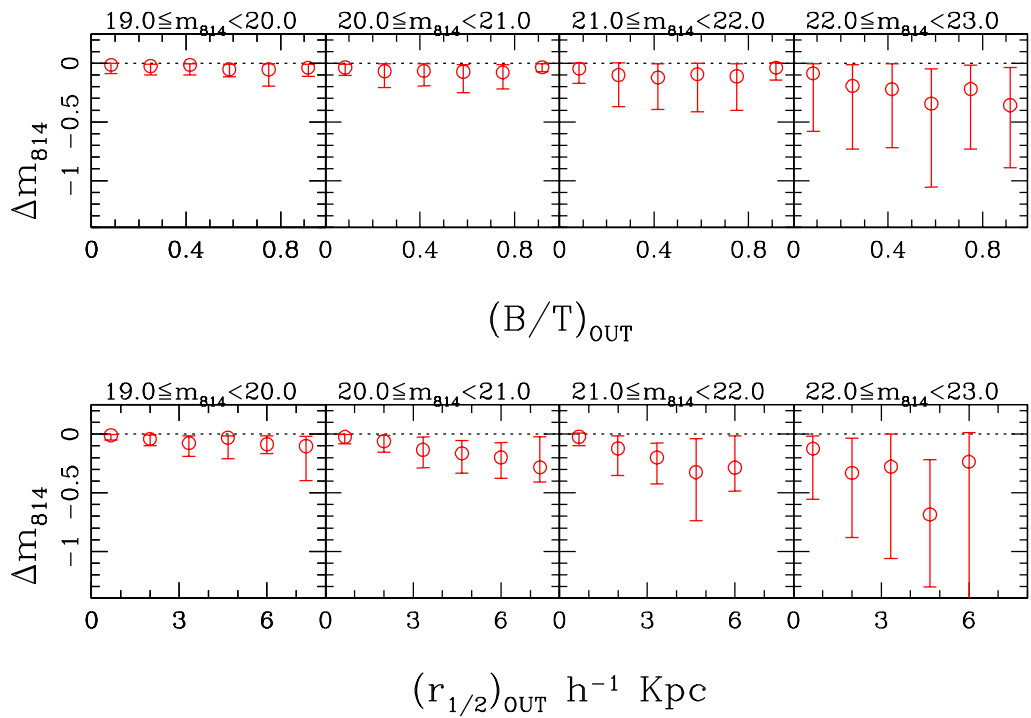}
\caption{{\it Top Panel:} Difference between input and measured
  magnitude ($\Delta m_{814}=IN-OUT$) as a function of measured $B/T$
  for a de Vaucouleurs bulge+exponential disk in the F814W filter
  (3600s integration).  The sample is split into four apparent
  (measured) magnitude bins with the brightest galaxies in the
  left-most subpanel.  Median values (open circles) and asymmetric
  errorbars corresponding to $1\sigma$ in the $\Delta m_{814}$
  distributions are shown. {\it Bottom Panel:} Same as in the top panel
  except now compared to the measured half-light radius (\rh)$_{OUT}$.
\label{1358flux_rh}}
\end{figure}

The top panel of Fig.~\ref{1358flux_rh} shows the median fractional
difference between input and measured magnitude (F814) over the range
of $B/T$ for the artificial catalog.  For the brightest galaxies, the
measured flux does not deviate significantly from the input value.  At
$m_{814}>22$~mags, however, we systematically underestimate the input
flux by $\sim0.3$ mags and maximum random errors increase to
$\Delta m_{814}>0.5$ across the range of $B/T$.  With the exception of
three galaxies (of 173), all of the CL~1358+62 sample are brighter than
$m_{814}=22$~mags.

The bottom panel of Fig.~\ref{1358flux_rh} emphasizes how faint
galaxies, and in particular low surface brightness ones, are adversely
affected by decreasing signal-to-noise ratios.  In each subpanel, we
underestimate the total flux as measured \rh~increases, $i.e.$ as
the average surface brightness decreases.  Even at fairly bright
magnitudes ($m_{814}\leq21$~mags), flux loss can be significant for
galaxies with \rh$>5$\hi kpc.  Fortunately, the median half-light
radius of the CL~1358+62 cluster sample is $2.4\pm1.2$\hi kpc so this
should not affect our conclusions.

\subsubsection{Bulge/Total Luminosity}

\begin{figure}
\plotone{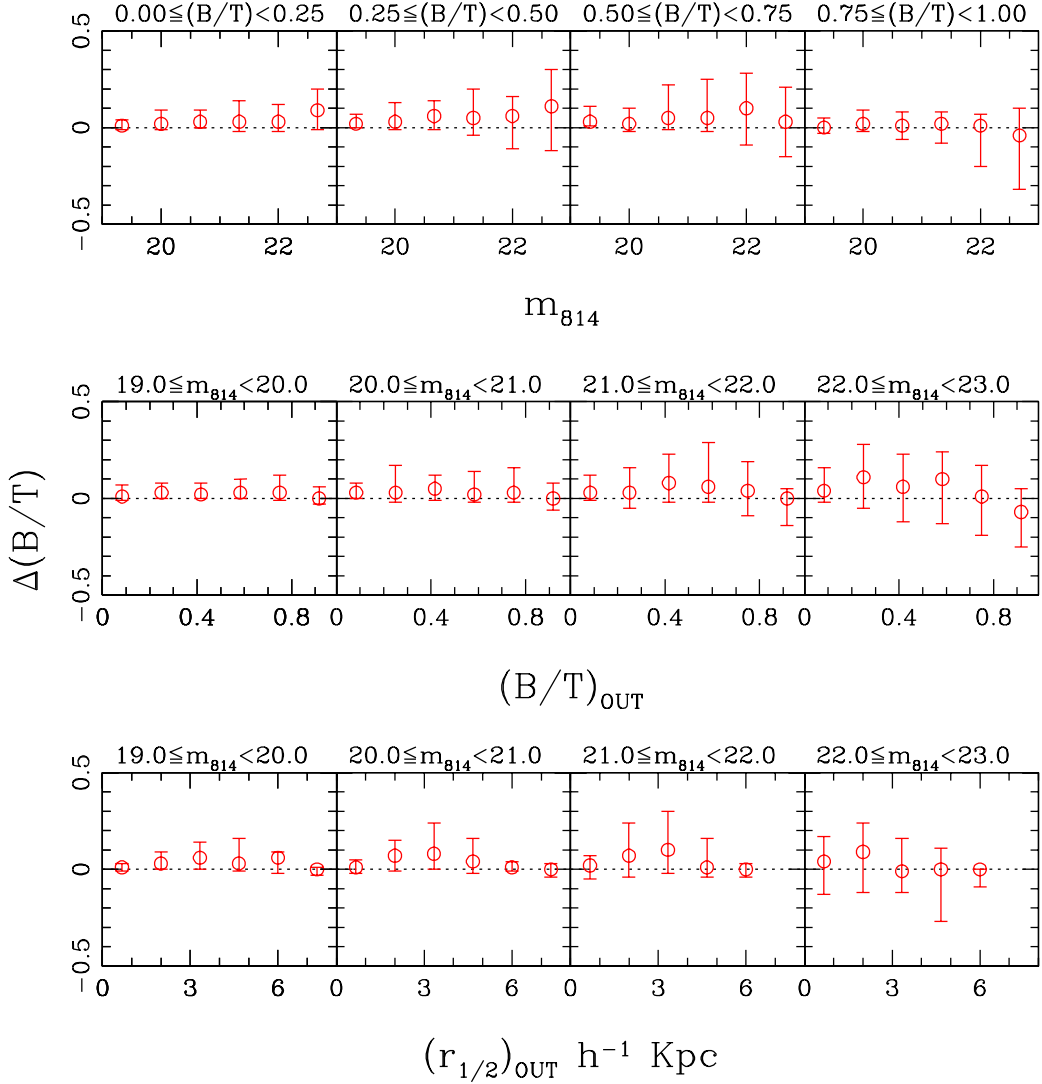}
\caption{Comparison of measured versus input values for bulge/total
  luminosity for a de Vaucouleurs bulge+exponential disk from the
  artificial galaxy catalog in the F814W filter (3600s integration).
  The y-axis shows $\Delta (B/T)=(B/T)_{IN}-(B/T)_{OUT}$.  {\it Top
    Panel:} The x-axis is measured F814W magnitude, and the sample is
  split into four $(B/T)_{OUT}$ bins with the most disk-dominated
  systems in the left-most subpanel.  {\it Middle Panel:} The x-axis is
  measured $B/T$ value, and the sample is split into four bins with the
  brightest galaxies in the left-most subpanel.  {\it Bottom Panel:}
  The x-axis is (\rh)$_{OUT}$, cut at \rh$=8$\hi Kpc to improve the
  statistics, and the sample is split into four bins where the
  brightest galaxies are in the left-most subpanel.  In all three
  panels, median values (open circles) and asymmetric errorbars
  corresponding to $1\sigma$ in the $\Delta (B/T)$ distributions are
  shown.  At magnitudes brighter than our cut of $m_{814}\leq21$,
  systematic errors are negligible [$\Delta (B/T)\sim0$] and random
  errors $<0.2$.
\label{1358bt_comp4}}
\end{figure}

\begin{figure}
\plotone{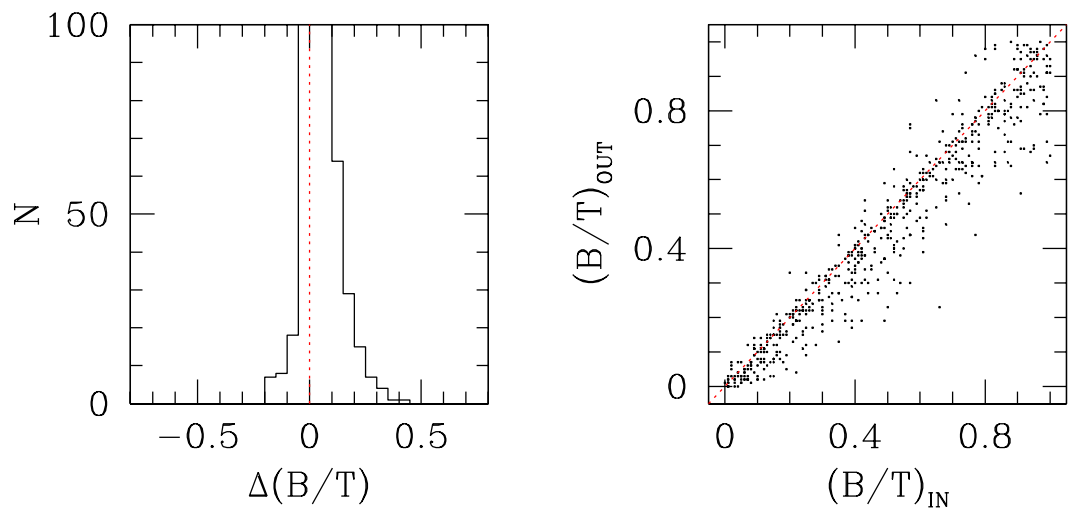}
\caption{{\it Left:} Histogram showing 
  [$\Delta (B/T)_{deV}=(B/T)_{IN}-(B/T)_{OUT}$]
  versus measured $(B/T)_{deV}$ for artificial galaxies brighter
  than $m_{814}=21$.  {\it Right:} Comparison of input to measured
  $(B/T)_{deV}$ for the same galaxies.
\label{1358dbt_deV}}
\end{figure}

Figure~\ref{1358bt_comp4} illustrates how the measured bulge fraction
[$(B/T)_{OUT}$] differs from its input value as a function of measured
$m_{814}$, $B/T$, and \rh.  Above our magnitude cut of $m_{814}=21$,
the systematic differences between input and measured $B/T$ are
negligible and the random errors are $<0.2$ (see
Fig.~\ref{1358dbt_deV}).  These results lend confidence to measurement
of the bulge/total luminosity for galaxies in our cluster sample.

\subsubsection{Half-Light Radius and Bulge/Disk Scale Lengths}

\begin{figure}
\plotone{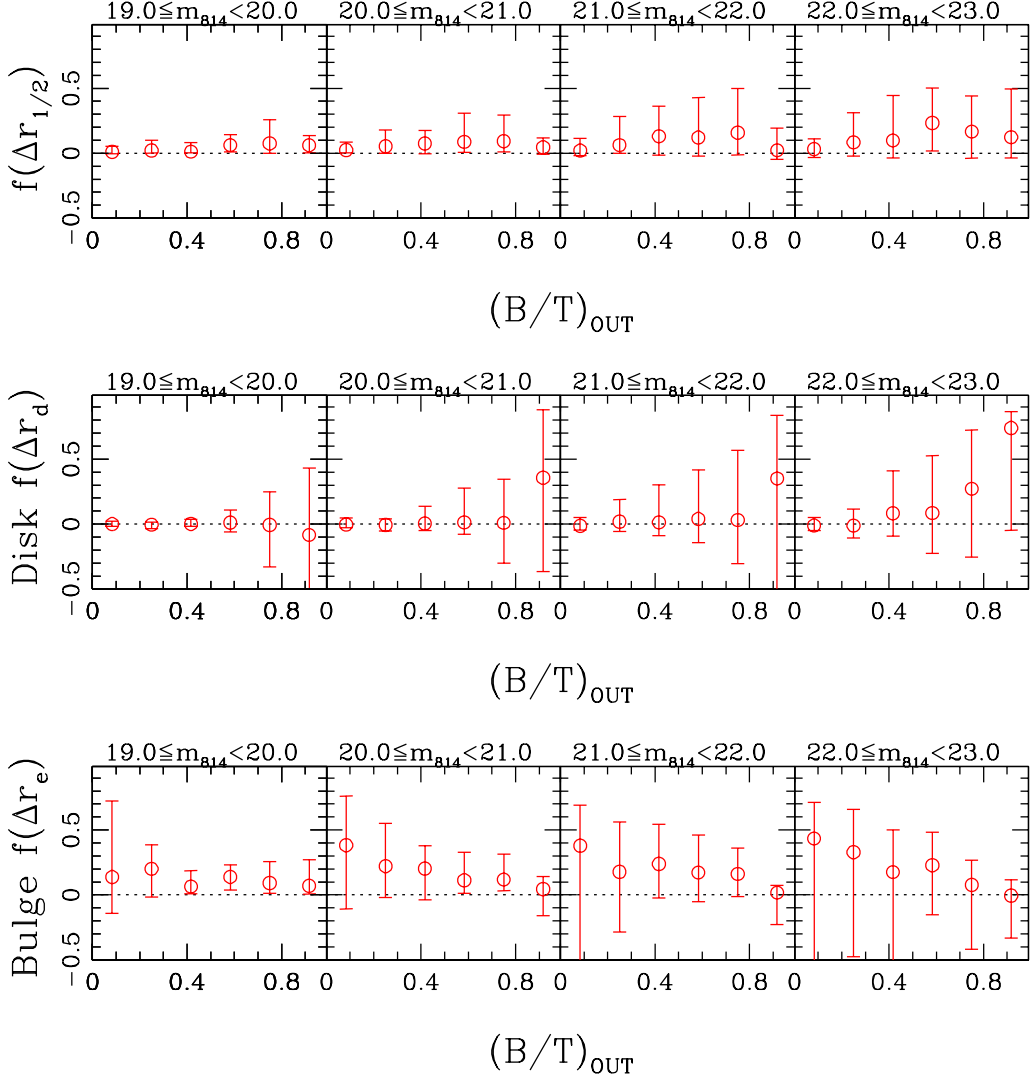}
\caption{For de Vaucouleurs bulge+exponential disk artificial
galaxies, the {\it fractional difference} between measured and input
values [$(IN-OUT)/OUT$] over the range in $(B/T)_{OUT}$ for \rh,
$r_d$, and $r_e$ (top, middle, and bottom panels respectively).
Asymmetric errorbars corresponding to $1\sigma$ in the distributions
are shown for each panel; lack of an errorbar indicates clumping at
that value.  Recovery of half-light radius \rh~is robust for
$(m_{814})_{OUT}\leq21$ galaxies with systematic and random errors of
$\sim10\%$ and $\sim20\%$ respectively.  The systematic errors
associated with disk scale lengths for $(B/T)_{OUT}<0.4$ galaxies are
negligible ($\sim0$) but random errors increase to $\sim50$\% at
$m_{814}>22$.  Comparatively, the bulge scale length $r_e$ tends to be
underestimated at all magnitudes and the random errors considerably
larger, especially for diskier systems ($B/T<0.6$).
\label{1358r_comp4}}
\end{figure}

\begin{figure}
\plotone{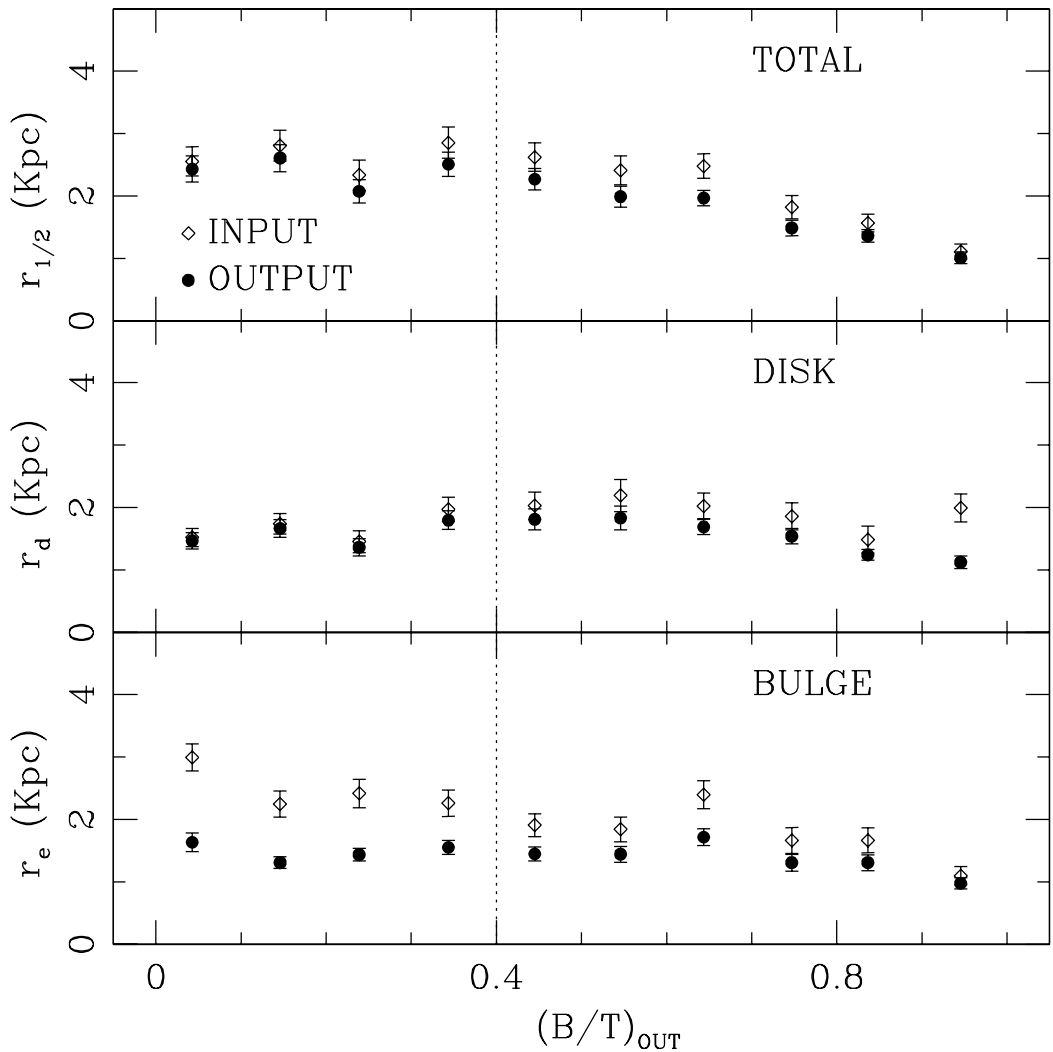}
\caption{Distribution of input (open symbols) and measured (filled
  symbols) values for half-light radii (top), disk (middle) and bulge
  (bottom panel) scale lengths as a function of measured $(B/T)_{deV}$
  (bin size 0.1).  We include only artificial galaxies with
  $(m_{814})_{OUT}\leq21$~mags, and $1\sigma$ errorbars are shown.  The
  dotted line denotes the adopted cut between bulge and disk-dominated
  systems.  Recovery of $r_d$ for $(B/T)_{OUT}<0.8$ galaxies is
  excellent.  Recovery of $r_e$ for disk-dominated systems is not as
  good, however, as due to the long wings of the de Vaucouleurs
  profile, flux at large radii can be buried in the background noise.
\label{1358rscales_bt}}
\end{figure}

Figure~\ref{1358r_comp4} (top panel) shows the fractional difference
between input and measured values of \rh~as a function of bulge
fraction.  Recovery of the half-light radius is robust for
$m_{814}\leq21$~mags galaxies with systematic and random errors of less
than 10\% and 20\% respectively.  This is reiterated in the top panel
of Fig.~\ref{1358rscales_bt}.  Here we show the average input and
measured \rh~values as a function of input bulge fraction.  Only for
the most bulge-dominated systems ($B/T\geq0.7$) are the differences
systematically larger than $1\sigma$ as due to the long wings of the de
Vaucouleurs profile, flux at large radii can be lost in the sky
background.

Recovery of disk exponential scale length for $B/T<0.4$ galaxies is
excellent.  At $m_{814}\leq21$~mags, systematic and random errors for
disk-dominated objects ($B/T<0.4$) are zero and $<15\%$ respectively
(middle panel, Fig.~\ref{1358r_comp4}).  In fact, the simulations show
that $r_d$ measurements are robust up to $B/T\sim0.8$ (middle panel,
Fig.~\ref{1358rscales_bt}).  

In comparison, measurements of the bulge scale length are not as
robust: $r_e$ tends to be underestimated ($\sim10-20$\%) at all
magnitudes (bottom panel, Figs.~\ref{1358r_comp4} \&
\ref{1358rscales_bt}).  Random errors, even for the most
bulge-dominated systems, also are larger than those associated with the
disk scale length.  While this result is not surprising when
considering the shape of the de Vaucouleurs profile, it can introduce a
bias that would be particularly detrimental to large, bulge-dominated,
low luminosity systems.

From these simulations, we determine that for the CL~1358+62 cluster
sample: 1) Measurements of half-light radius across the range of $B/T$
are robust; 2) Disk scale lengths are reliable for galaxies with
substantial disks ($B/T<0.6$); and 3) The systematic uncertainties
associated with measuring bulge scale lengths requires that care be
applied in their interpretation, especially for galaxies with a small
bulge component ($B/T<0.4$).

\subsubsection{Bulge Ellipticity \& Disk Inclination}

\begin{figure}
\plotone{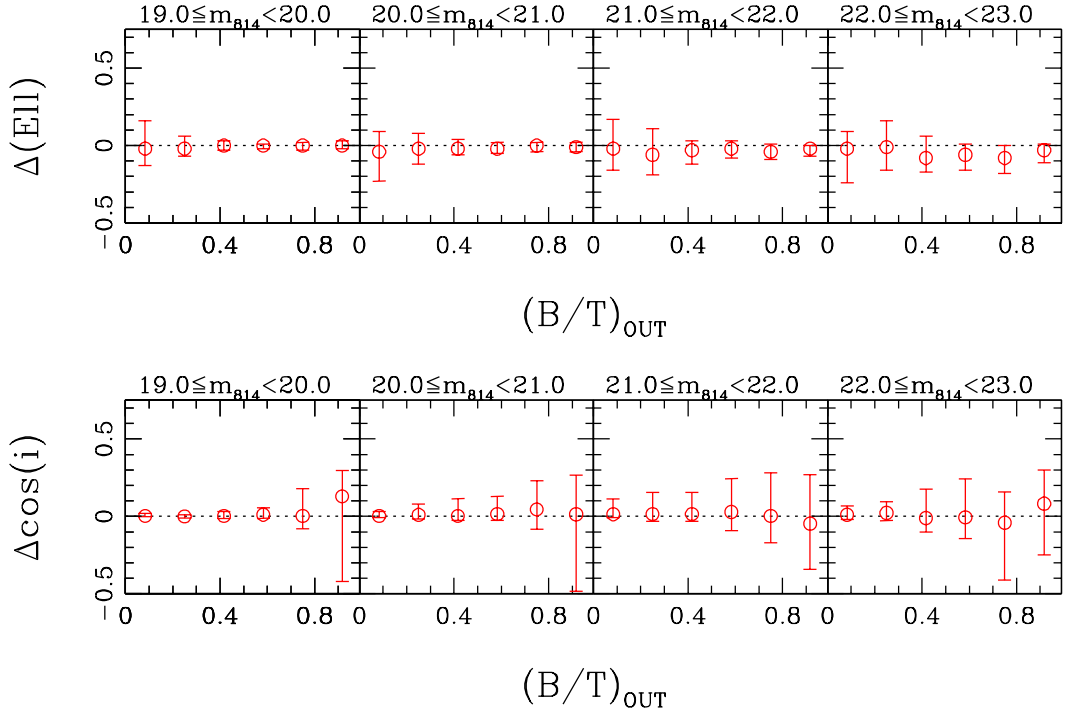}
\caption{{\it Top Panel:} The difference between input and measured
  bulge ellipticity ($IN-OUT$) for the artificial galaxies (de
  Vaucouleurs bulge+exponential disk).  Recovery of bulge ellipticity
  for these objects is excellent, except for the most disk-dominated
  systems.  The average difference is zero, and the associated random
  $1\sigma$ error is $\Delta (ell)<0.1$.  {\it Bottom Panel:} The
  difference between input and measured disk inclination for the
  artificial galaxies.  Here, a face-on disk has cos~$i=1$.  The
  recovery of disk inclination for these galaxies also is robust with
  an systematic difference of zero, except for the most bulge-dominated
  systems; the associated random error is $\Delta$(cos~$i)<0.1$.
\label{1358cosi_ell4}}
\end{figure}

To test the robustness of our results concerning ellipticals and S0's,
we compare input and measured values of bulge ellipticity and disk
inclination.  Figure~\ref{1358cosi_ell4} shows the average differences
in bulge ellipticity and disk inclinations for artificial galaxies as
a function of measured $B/T$ and apparent magnitude.  At
$m_{814}\leq21$, both parameters are recovered well with median
differences between input and measured values of approximately zero.
The associated random errors in both ellipticity and cos~$i$ are
$<0.1$.  As expected, random errors in the bulge ellipticity are
largest for disk-dominated systems, and errors in cos~$i$ largest in
bulge-dominated systems.

\subsection{Determining S\'ersic Bulges}

We find that fitting S\'ersic bulges with exponential disks to the
cluster members results in a significant correlation between visual
type and bulge fraction (see Fig.~\ref{1358profiles}).  By fitting
S\'ersic profiles, however, we must test our ability to recover the
true bulge power $n$.  To do so, we create an artificial galaxy
catalog of $1500$ galaxies in the F814W filter with bulge fraction
$\geq0.5$ and bulge power $n$ between 0.2 and 6; all other structural
parameters are the same as for the deVaucouleurs bulge+exponential
disk artificial catalog.  As before, we discuss results only for
galaxies with $m_{814}\leq21$~mags.

\subsubsection{Recovery of the S\'ersic Index $n$}

\begin{figure}
\plotone{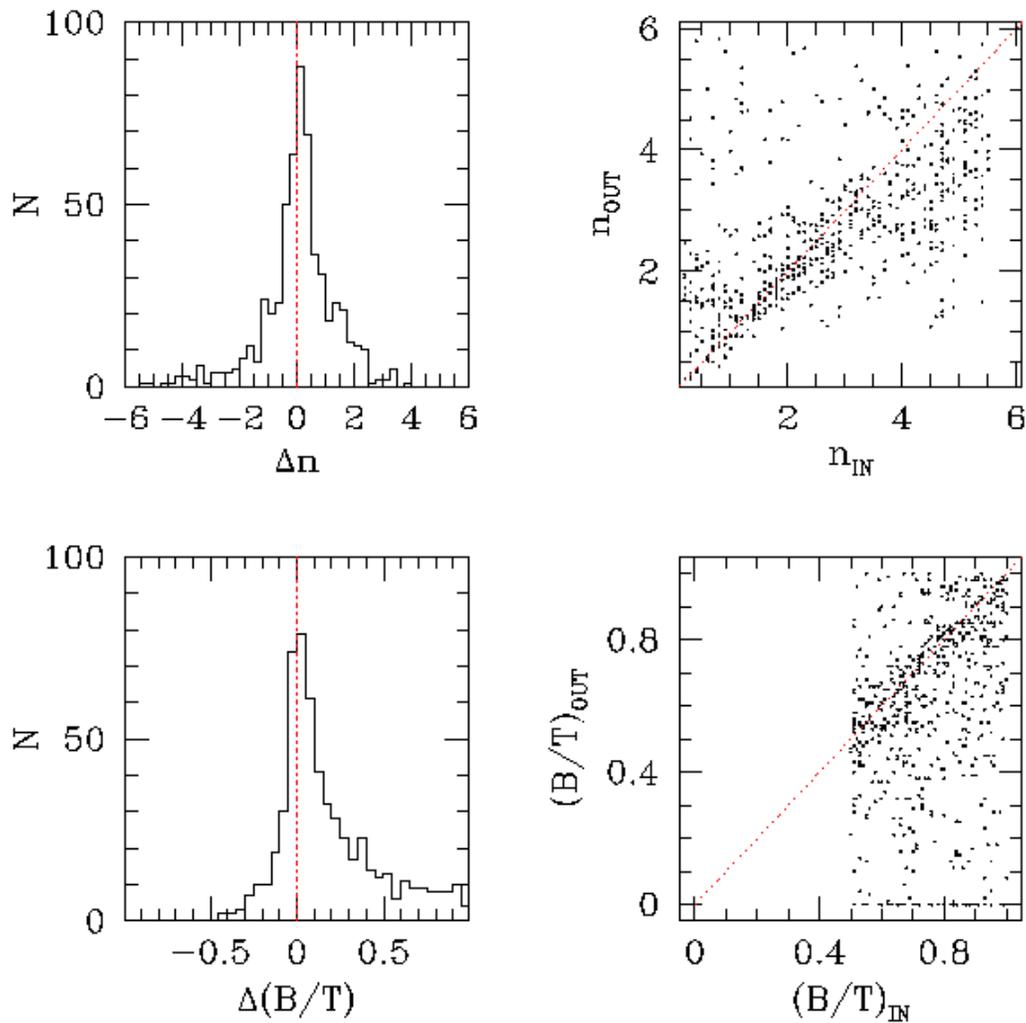}
\caption{{\it Top Panels:} The difference between input versus
  measured value ($IN-OUT$) of the bulge power $n$ ($0.2\leq n\leq6$)
  from the artificial S\'ersic catalog of $(B/T)_{IN}\geq0.5$ galaxies;
  only galaxies with $(m_{814})_{OUT}\leq21$~mags are included.
  Although 70\% of the galaxies have differences $\leq0.5$, $\Delta n$
  can range up to $\sim5$ for some.  {\it Bottom Panel:} The difference
  in bulge fraction for the same galaxies.  Here, the distribution is
  skewed towards underestimating the true bulge fraction where the
  median difference is $\sim0.14$.  Many galaxies ($\sim25$\%) now are
  considered disk-dominated systems [$(B/T)_{OUT}\leq0.4$] although
  none have $(B/T)_{IN}<0.5$.
\label{1358dnb_plot}}
\end{figure}

For most galaxies (70\%), we are within 0.5 of the true S\'ersic bulge
power $n$ (Fig.~\ref{1358dnb_plot}; top panel).  Approximately 20\% of the
catalog, however, has $\Delta n\geq1$ and, in some cases, $\Delta n$
can be as much as $\sim4$.  Even more disconcerting is the comparison
of input versus measured bulge fraction (Fig.~\ref{1358dnb_plot}; lower
panel).  The distribution is skewed heavily towards underestimating
the true bulge fraction (median $\Delta (B/T)\sim14\%$), and 23\% of
the galaxies are now considered disk-dominated systems even though
none have input $B/T<0.5$.

These simulations emphasize that measuring reliable S\'ersic $n$ and
the corresponding bulge fraction requires 1) a combination of excellent
resolution and high signal-to-noise; 2) understanding how noise affects
the measurements; and 3) rigorous testing of structural parameters
derived for each galaxy.

\subsubsection{Fitting S\'ersic Bulges with $r^{1/4}$ Profiles}

\begin{figure}
\plotone{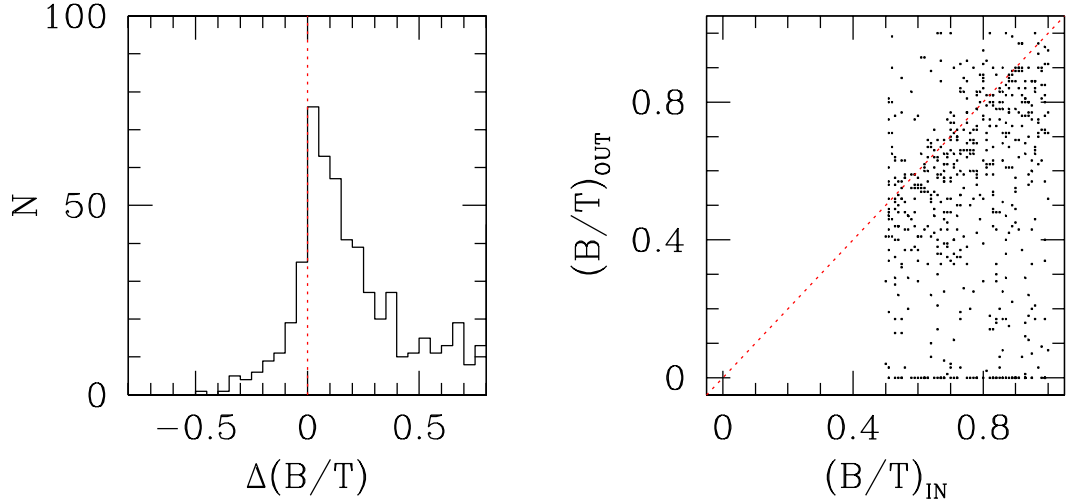}
\caption{{\it Left:} Histogram showing the difference between input
  versus measured value ($IN-OUT$) of $B/T$ for {\it S\'ersic bulges
    fitted using the $r^{1/4}$ profile}; all of the artificial galaxies
  have $(B/T)_{IN}\geq0.5$.  Here we only consider galaxies with
  measured $m_{814}$ brighter than $21$.  {\it Right:} Comparison of
  input to measured $B/T$ for the same galaxies.  As in the case of the
  S\'ersic bulge fitted with the S\'ersic profile, the true bulge
  fraction tends to be underestimated when modeling a S\'ersic bulge
  with an $r^{1/4}$ profile.
\label{1358dbt_cross}}
\end{figure}

With the artificial catalog of S\'ersic bulge+exponential disk
galaxies, we address the question of how serious the errors would be
if we modeled them as {\it de Vaucouleurs ($r^{1/4})$ bulges}.
Figure~\ref{1358dbt_cross} compares the input S\'ersic $B/T$ compared
to that recovered using the de Vaucouleurs profile.  The distribution
is as skewed towards underestimating the true bulge fraction as in
Fig.~\ref{1358dnb_plot}; this is due primarily to high $B/T$ systems
with bulge $n\sim1$ being confused as disk-dominated systems (see \S5.3).  
Even for galaxies with higher S\'ersic indices ($n>3$), the 
average $\Delta (B/T)$ is still larger than if the true bulge profile
is $r^{1/4}$.

However, we find that the {\it half-light radii} ($\sim$sizes) are
robust.  As for bulge ellipticities and bulge scale lengths of {\it
bulge-dominated} [$(B/T)_{OUT}\geq0.4$] galaxies, the systematic
errors are still negligible but the random errors are larger by
$\sim2$ and $\sim3$ respectively.  Despite the uncertainty in $(B/T)$,
certain structural parameters such as half-light radius as well as
bulge ellipticity and scale length [for $(B/T)_{OUT}\geq0.4$] are
reliable.

\subsection{Determining Exponential Bulges}

Although the correlation between visual type and bulge fraction as
determined using a double exponential profile is poor (see
Fig.\ref{1358profiles}), we test our ability to recover this profile
for completeness.  We create an artificial galaxy catalog of 1500
galaxies with exponential bulges and disks in the F814W filter that
spans the range of $B/T$.  Again, we limit our discussion to galaxies
with $m_{814}\leq21$~mags.

\begin{figure}
\plotone{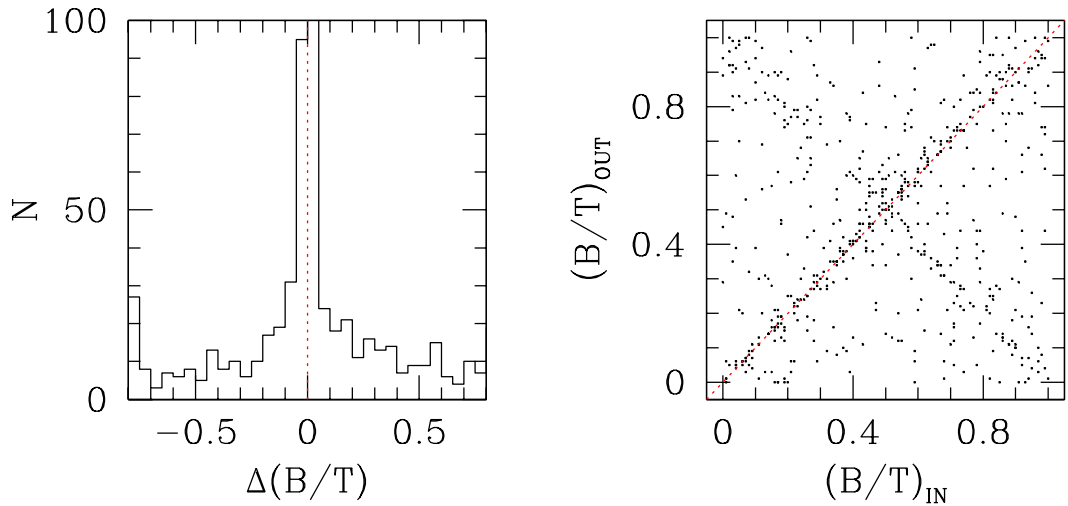}
\caption{{\it Left:} Histogram showing the difference between input
  versus measured value ($IN-OUT$) of $B/T$ for the artificial double
  exponential catalog; only galaxies with $(m_{814})_{OUT}\leq21$~mags
  are included.  {\it Right:} Comparison of input to measured $B/T$ for
  the same galaxies.  The obvious problem with using a double
  exponential profile is that the bulge/disk components are often
  reversed, i.e. bulge-dominated systems become disk-dominated ones and
  vice versa.
\label{1358ee_dbt}}
\end{figure}

Figure ~\ref{1358ee_dbt} illustrates how the bulge/disk components
often are reversed such that bulge-dominated systems become
disk-dominated ones and vice versa.  This ambiguity manifests as the
long tails in the two panels of Fig.~\ref{1358ee_dbt}.  While our
inability to recover the true double exponential profile is not a
compelling argument for not using it, these simulations demonstrate
that a more sophisticated version of this profile is needed.  For
example, incorporating a truncation radius
\citep{vanderkruit:82,vanderkruit:87,degrijs:01,kregel:02} for the
bulge component may alleviate the ambiguity.


\section{Conclusions}

Combining deep wide field HST imaging ($8'\times8'$ field;
$2.2\times2.2$ Mpc$^2$) with ground-based spectroscopy, we characterize
the physical properties of 173 cluster members in CL~1358+62
($z=0.3283$).  Our study is unique due to the large number of confirmed
cluster members, the distance to the cluster, and the high physical
resolution of the galaxies ($0.27$\hi~kpc).  By fitting two component
surface profiles directly to each galaxy, we measure their structural
parameters and compare these results to current star formation, total
galaxy colors, and cluster substructure.  In this study, we examine the
viability of three different bulge profiles (de Vaucouleurs, S\'ersic,
and exponential) paired with an exponential disk component.

We stringently test the robustness of our results by fitting thousands
of artificial galaxies in the same manner as the sample; note that we
assume galaxies are well-described by an $r^{1/n}$ bulge+exponential
disk profile.  As the artificial galaxy catalog spans the same range in
luminosity, bulge fraction, half-light radius, and scale lengths as the
cluster sample, we can quantify the systematic and random errors
associated with the structural parameter measurements.  From these
simulations, we find that we tend to underestimate galaxy bulge
fraction, especially for bulge-dominated objects.  However, over the
selected magnitude range ($m_{814}\leq21$~mags; $MB_z<-17.3$\logh) for
the de Vaucouleurs bulge+exponential disk profile, measurements of
$B/T$ have systematic errors of $<15$\%.

Our conclusions are as follows:

\begin{enumerate}
\item In this sample, the correlation between $(B/T)_{deV}$ and Hubble
type ($-5\leq T\leq 10$) is strong (99\% confidence), but the scatter
is large.  In particular, early-type spirals cannot be distinguished
from ellipticals and S0's based on their bulge fraction.
\item Using quantitative morphological parameters, we find a 
predominance of bulge-dominated (early-type) galaxies and an established
morphology-density relation, both characteristics of a well-developped
cluster.
\item The physical properties, $i.e.$ bulge ellipticities, disk
inclinations, $B/T$, half-light radii, and internal velocity
dispersions, of the low luminosity ($-17.3\geq MB_z-5\log h\geq-19.3$)
ellipticals and S0's in our sample suggest that these ellipticals are
likely to be face-on S0 galaxies.
\item We find a population ($\sim5$\%) of bulge-dominated, massive
($\sigma>130$\kms) members with significant [OII]$\lambda3727$
emission ($<-5$\AA), morphologically smooth profiles, and colors as 
red as the absorption galaxies; the latter is true for both their total
and bulge component.  These physical characteristics suggest they
harbor AGN.
\item The most disk-dominated members ($B/T<0.25$) have significantly
higher average [OII]$\lambda3727$ equivalent widths than members with
$B/T\geq0.25$.  Disk-dominated galaxies also tend to have high degrees
of structural asymmetry.
\item Mimicking the sample with thousands of artificial galaxies is
integral in understanding how robust correlations between physical and
spectral properties are.  In particular, measuring accurate S\'ersic
indices at these redshifts requires significantly higher
signal-to-noise and/or resolution than that of our WFPC2 imaging.
\end{enumerate}

\acknowledgments

Sincere thanks to D. Kelson and P. van Dokkum for their advice
throughout this work.  We appreciate the useful comments S. Faber and
the referee provided on this manuscript. Support from NASA HST grants
GO-05989.01 and GO-07372.01, and NASA grant NAG5-7697 is gratefully
acknowledged.

\bibliographystyle{/home/vy/aastex/apj}
\bibliography{/home/vy/aastex/tran.bib}

\clearpage

\clearpage

\clearpage

\clearpage


\begin{deluxetable}{rrrrrrrrrr}
\tablecolumns{10}
\tablewidth{0pc}
\tablecaption{CL~1358+62 Structural Parameters\label{params}}
\tablehead{
\colhead{ID}    & \colhead{Type\tablenotemark{a}}       &
\colhead{$(B/T)$\tablenotemark{b}}      &
\colhead{$B_z$\tablenotemark{c}}        &
\colhead{$(B-V)_z$\tablenotemark{c}}    &
\colhead{$(r_{1/2}'')$\tablenotemark{b}}        &
\colhead{$(r_e '')$\tablenotemark{b}}   &
\colhead{$(r_d '')$\tablenotemark{b}}   &
\colhead{$(R_A)$\tablenotemark{b}}      & \colhead{$(R_T)$\tablenotemark{b}}}
\startdata
92 & --2 & 0.75 & 21.49 & 0.85 & 0.47 & 0.51 & 0.25 & 0.02 & 0.09 \\
95 & --2 & 0.54 & 21.02 & 0.89 & 0.47 & 0.19 & 0.55 & 0.02 & 0.06 \\
97 & 1 & 0.24 & 21.06 & 0.82 & 1.17 & 0.49 & 0.81 & 0.04 & 0.10 \\
105 & --4 & 0.52 & 22.26 & 0.87 & 0.34 & 0.11 & 0.43 & 0.01 & 0.05 \\
108 & --4 & 0.81 & 21.22 & 0.87 & 0.76 & 0.55 & 1.25 & 0.01 & --0.17 \\
109 & --2 & 0.41 & 21.95 & 0.67 & 0.46 & 0.16 & 0.41 & 0.01 & 0.07 \\
110 & 0 & 0.40 & 20.56 & 0.85 & 0.77 & 0.28 & 0.66 & 0.01 & 0.06 \\
126 & --2 & 0.38 & 20.71 & 0.85 & 0.66 & 0.16 & 0.62 & 0.01 & 0.06 \\
129 & 0 & 0.90 & 20.66 & 0.83 & 0.41 & 0.34 & 1.39 & 0.02 & 0.06 \\
132 & 3 & 0.05 & 22.58 & 0.64 & 0.72 & 0.60 & 0.43 & 0.05 & 0.12 \\
135 & --2 & 0.41 & 21.40 & 0.83 & 0.41 & 0.10 & 0.41 & 0.02 & 0.06 \\
137 & 3 & 0.00 & 22.04 & 0.80 & 0.54 & \nodata & 0.32 & 0.05 & 0.14 \\
142 & 0 & 0.31 & 20.82 & 0.86 & 0.96 & 0.40 & 0.71 & 0.01 & 0.07 \\
143 & --4 & 0.60 & 21.73 & 0.87 & 0.38 & 0.29 & 0.29 & 0.01 & 0.04 \\
149 & --4 & 0.80 & 21.59 & 0.84 & 0.56 & 0.45 & 0.56 & 0.01 & 0.06 \\
164 & 0 & 0.42 & 20.36 & 0.87 & 1.19 & 0.45 & 1.05 & 0.03 & 0.10 \\
167 & --2 & 0.72 & 22.07 & 0.60 & 0.53 & 0.38 & 0.54 & 0.02 & 0.08 \\
171 & 0 & 0.42 & 22.47 & 0.74 & 0.48 & 0.13 & 0.48 & 0.02 & 0.09 \\
178 & 5 & 0.00 & 20.98 & 0.50 & 1.06 & \nodata & 0.63 & 0.13 & 0.18 \\
182 & --2 & 0.71 & 21.77 & 0.84 & 0.41 & 0.32 & 0.36 & 0.01 & 0.04 \\
190 & 4 & 0.09 & 22.42 & 0.82 & 0.93 & 0.65 & 0.56 & 0.03 & 0.10 \\
192 & 6 & 0.00 & 21.71 & 0.53 & 1.15 & \nodata & 0.68 & 0.06 & 0.11 \\
200 & 0 & 0.00 & 20.45 & 0.71 & 1.42 & \nodata & 0.85 & 0.05 & 0.15 \\
203 & --2 & 0.37 & 21.41 & 0.84 & 0.92 & 0.33 & 0.75 & 0.02 & 0.07 \\
205 & --2 & 0.52 & 22.57 & 0.80 & 0.23 & 0.07 & 0.32 & 0.02 & 0.07 \\
206 & 0 & 0.03 & 22.12 & 0.77 & 0.83 & 0.50 & 0.50 & 0.01 & 0.10 \\
207 & --2 & 0.13 & 22.48 & 0.80 & 0.72 & 0.12 & 0.48 & 0.00 & 0.05 \\
209 & 1 & 0.42 & 20.47 & 0.60 & 0.99 & 0.87 & 0.62 & 0.03 & 0.08 \\
210 & --5 & 0.89 & 21.05 & 0.89 & 0.55 & 0.55 & 0.34 & 0.02 & 0.06 \\
211 & --2 & 0.48 & 20.82 & 0.87 & 0.71 & 0.30 & 0.67 & 0.03 & 0.07 \\
212 & --5 & 0.79 & 20.65 & 0.87 & 0.71 & 0.48 & 1.43 & 0.01 & 0.07 \\
215 & --2 & 0.57 & 21.54 & 0.84 & 0.37 & 0.16 & 0.44 & 0.02 & 0.06 \\
226 & 1 & 0.27 & 21.79 & 0.68 & 0.86 & 0.60 & 0.55 & 0.02 & 0.08 \\
227 & 0 & 0.38 & 21.73 & 0.91 & 0.73 & 0.40 & 0.53 & 0.04 & 0.08 \\
233 & --4 & 0.84 & 19.97 & 0.85 & 0.95 & 0.71 & 2.03 & 0.04 & 0.09 \\
234 & 3 & 0.09 & 20.17 & 0.69 & 1.66 & 0.08 & 1.09 & 0.10 & 0.17 \\
235 & --2 & 0.91 & 22.33 & 0.81 & 0.70 & 0.74 & 0.32 & 0.03 & 0.13 \\
236 & --2 & 0.62 & 20.83 & 0.89 & 0.74 & 0.36 & 1.01 & 0.01 & 0.06 \\
239 & --4 & 0.53 & 21.48 & 0.90 & 0.33 & 0.15 & 0.33 & 0.02 & 0.06 \\
242 & --5 & 1.00 & 20.21 & 0.88 & 1.45 & 1.45 & \nodata & 0.02 & 0.13 \\
243 & 1 & 0.68 & 21.99 & 0.41 & 0.48 & 0.55 & 0.25 & 0.05 & --0.02 \\
246 & 1 & 0.46 & 21.72 & 0.84 & 0.65 & 0.36 & 0.52 & 0.01 & 0.09 \\
248 & --2 & 0.68 & 21.15 & 0.93 & 0.32 & 0.18 & 0.44 & 0.01 & 0.04 \\
254 & --4 & 0.61 & 20.65 & 0.87 & 0.68 & 0.55 & 0.49 & 0.01 & 0.04 \\
256 & --5 & 0.70 & 19.36 & 0.88 & 1.20 & 0.74 & 1.49 & 0.01 & 0.04 \\
264 & --99 & 0.00 & 22.41 & 0.06 & 0.85 & \nodata & 0.51 & 0.10 & 0.17 \\
265 & 0 & 0.67 & 21.28 & 0.88 & 0.60 & 0.41 & 0.55 & 0.02 & --0.03 \\
269 & --4 & 0.79 & 19.61 & 0.91 & 1.07 & 0.81 & 1.29 & 0.01 & 0.04 \\
271 & --99 & 0.78 & 22.36 & 0.40 & 0.24 & 0.22 & 0.18 & 0.02 & 0.06 \\
272 & --5 & 0.83 & 21.77 & 0.89 & 0.29 & 0.23 & 0.36 & 0.01 & 0.05 \\
278 & --2 & 0.82 & 21.42 & 0.88 & 0.39 & 0.41 & 0.21 & 0.02 & 0.07 \\
279 & --5 & 0.86 & 21.32 & 0.80 & 0.49 & 0.41 & 0.54 & 0.02 & 0.05 \\
285 & 1 & 0.52 & 22.47 & 0.87 & 0.41 & 0.24 & 0.34 & 0.02 & 0.07 \\
288 & --2 & 0.61 & 21.53 & 0.87 & 0.38 & 0.21 & 0.41 & 0.01 & --0.06 \\
290 & 6 & 0.00 & 21.70 & 0.61 & 0.71 & \nodata & 0.42 & 0.07 & 0.12 \\
292 & 0 & 0.29 & 20.76 & 0.78 & 0.57 & 0.33 & 0.39 & 0.03 & 0.07 \\
293 & --2 & 0.64 & 22.43 & 0.85 & 0.25 & 0.10 & 0.57 & 0.03 & 0.09 \\
295 & --2 & 0.86 & 20.89 & 0.86 & 0.72 & 0.71 & 0.47 & 0.02 & 0.06 \\
298 & --2 & 0.68 & 20.20 & 0.91 & 0.78 & 0.93 & 0.38 & 0.02 & 0.05 \\
299 & --5 & 0.45 & 21.98 & 0.83 & 0.61 & 0.16 & 0.67 & 0.03 & 0.09 \\
300 & --2 & 0.77 & 20.85 & 0.88 & 0.40 & 0.31 & 0.39 & 0.01 & 0.04 \\
303 & --5 & 0.84 & 20.70 & 0.86 & 0.62 & 0.47 & 1.15 & 0.02 & 0.05 \\
305 & 3 & 0.24 & 21.56 & 0.83 & 0.90 & 0.29 & 0.65 & 0.05 & 0.05 \\
309 & --4 & 0.79 & 20.58 & 0.90 & 0.54 & 0.39 & 0.71 & 0.02 & 0.05 \\
311 & --2 & 0.65 & 22.14 & 0.84 & 0.25 & 0.15 & 0.28 & 0.02 & 0.05 \\
318 & --2 & 0.68 & 21.75 & 0.81 & 0.44 & 0.27 & 0.51 & 0.02 & 0.08 \\
323 & --2 & 0.37 & 21.91 & 0.84 & 0.44 & 0.14 & 0.38 & 0.02 & 0.07 \\
328 & 1 & 0.45 & 20.62 & 0.63 & 0.68 & 0.47 & 0.47 & 0.05 & 0.09 \\
329 & --2 & 0.66 & 21.43 & 0.82 & 0.77 & 0.59 & 0.61 & 0.02 & 0.09 \\
331 & --99 & 0.84 & 22.77 & 0.86 & 0.11 & 0.08 & 0.38 & 0.01 & 0.06 \\
335 & 0 & 0.62 & 21.28 & 0.86 & 0.62 & 0.37 & 0.60 & 0.02 & 0.09 \\
343 & --2 & 0.91 & 21.35 & 0.67 & 0.50 & 0.52 & 0.25 & 0.01 & 0.06 \\
344 & 0 & 0.38 & 21.42 & 0.83 & 0.94 & 0.63 & 0.63 & 0.01 & 0.07 \\
346 & --2 & 0.62 & 20.53 & 0.78 & 0.53 & 0.25 & 0.70 & 0.02 & 0.05 \\
347 & --5 & 0.51 & 20.61 & 0.86 & 0.98 & 0.34 & 1.15 & 0.01 & 0.01 \\
349 & 0 & 0.50 & 22.21 & 0.83 & 0.69 & 0.32 & 0.63 & 0.01 & 0.06 \\
350 & --2 & 0.74 & 22.18 & 0.89 & 0.21 & 0.18 & 0.16 & 0.03 & 0.08 \\
351 & 3 & 0.50 & 20.67 & 0.76 & 0.86 & 0.52 & 0.67 & 0.07 & 0.13 \\
352 & --5 & 0.70 & 21.92 & 0.88 & 0.31 & 0.19 & 0.40 & 0.01 & 0.05 \\
354 & --5 & 0.92 & 22.03 & 0.87 & 0.20 & 0.19 & 0.13 & 0.02 & 0.04 \\
356 & 1 & 0.50 & 20.25 & 0.89 & 0.97 & 0.39 & 0.99 & 0.02 & 0.08 \\
357 & --5 & 0.78 & 21.39 & 0.87 & 0.21 & 0.13 & 0.70 & 0.04 & 0.19 \\
358 & --5 & 0.84 & 21.82 & 0.87 & 0.44 & 0.34 & 0.74 & 0.03 & 0.07 \\
359 & --2 & 0.87 & 20.55 & 0.85 & 0.56 & 0.44 & 1.28 & 0.02 & 0.08 \\
361 & 1 & 0.39 & 21.81 & 0.85 & 0.55 & 0.21 & 0.46 & 0.01 & 0.06 \\
366 & --4 & 0.71 & 21.01 & 0.86 & 0.57 & 0.38 & 0.63 & 0.01 & 0.07 \\
368 & 2 & 0.84 & 21.18 & 0.86 & 2.69 & 2.14 & 3.55 & --0.01 & 0.18 \\
369 & 0 & 0.67 & 21.06 & 0.88 & 0.86 & 0.61 & 0.77 & 0.02 & 0.07 \\
371 & 1 & 0.22 & 20.05 & 0.87 & 1.41 & 0.38 & 1.01 & 0.01 & 0.06 \\
375 & --5 & 0.48 & 19.20 & 0.89 & 2.90 & 1.25 & 2.73 & 0.02 & --0.01 \\
376 & --2 & 0.47 & 21.30 & 0.88 & 0.36 & 0.10 & 0.44 & 0.03 & 0.06 \\
377 & 5 & 0.05 & 20.43 & 0.64 & 1.84 & 1.32 & 1.10 & 0.06 & 0.16 \\
380 & 6 & 0.00 & 21.28 & 0.53 & 1.14 & \nodata & 0.68 & 0.06 & 0.15 \\
381 & --4 & 0.64 & 20.72 & 0.87 & 0.94 & 0.44 & 1.52 & 0.01 & 0.09 \\
383 & 999 & 0.40 & 22.08 & 0.84 & 0.58 & 0.33 & 0.43 & 0.03 & 0.07 \\
388 & 8 & 0.25 & 22.76 & 0.86 & 0.49 & 0.35 & 0.31 & 0.01 & 0.05 \\
394 & 0 & 0.21 & 22.07 & 0.84 & 0.51 & 0.07 & 0.39 & 0.02 & 0.02 \\
396 & 8 & 0.00 & 21.46 & 0.45 & 1.02 & \nodata & 0.61 & 0.11 & 0.18 \\
400 & --2 & 0.54 & 22.38 & 0.85 & 0.15 & 0.04 & 0.24 & 0.01 & 0.05 \\
406 & --5 & 0.81 & 22.57 & 0.76 & 0.38 & 0.33 & 0.34 & 0.00 & 0.07 \\
408 & --2 & 0.89 & 21.02 & 0.86 & 0.44 & 0.43 & 0.29 & 0.01 & 0.06 \\
409 & --5 & 0.61 & 21.90 & 0.90 & 0.49 & 0.27 & 0.50 & 0.01 & 0.06 \\
410 & --2 & 0.58 & 21.43 & 0.87 & 0.53 & 0.24 & 0.64 & 0.02 & 0.07 \\
411 & 0 & 0.50 & 21.80 & 0.80 & 0.58 & 0.38 & 0.44 & 0.01 & 0.07 \\
416 & --2 & 0.64 & 22.33 & 0.84 & 0.20 & 0.12 & 0.20 & 0.02 & 0.04 \\
420 & --99 & 0.45 & 22.71 & 0.62 & 0.52 & 0.35 & 0.37 & --0.01 & 0.04 \\
421 & --5 & 0.65 & 21.59 & 0.90 & 0.43 & 0.23 & 0.56 & 0.03 & 0.08 \\
434 & --5 & 0.79 & 22.50 & 0.88 & 0.12 & 0.07 & 0.32 & 0.02 & 0.04 \\
440 & 0 & 0.38 & 21.60 & 0.84 & 0.53 & 0.20 & 0.44 & 0.02 & 0.07 \\
442 & 0 & 0.71 & 23.23 & 0.85 & 0.24 & 0.18 & 0.21 & 0.02 & 0.07 \\
444 & --2 & 0.60 & 21.01 & 0.90 & 0.70 & 0.67 & 0.43 & 0.01 & 0.06 \\
447 & --2 & 0.44 & 21.58 & 0.90 & 0.52 & 0.21 & 0.46 & 0.02 & 0.08 \\
453 & 5 & 0.35 & 20.78 & 0.52 & 1.37 & 2.02 & 0.73 & 0.07 & 0.15 \\
454 & 0 & 0.51 & 20.55 & 0.81 & 0.95 & 1.41 & 0.46 & 0.02 & 0.07 \\
457 & 0 & 0.23 & 21.72 & 0.85 & 0.56 & 0.11 & 0.42 & 0.01 & 0.07 \\
460 & --2 & 0.36 & 22.26 & 0.84 & 0.37 & 0.11 & 0.32 & 0.01 & 0.05 \\
463 & --2 & 0.77 & 20.59 & 0.88 & 0.71 & 0.59 & 0.59 & 0.02 & 0.05 \\
465 & 1 & 0.64 & 21.26 & 0.87 & 0.63 & 0.41 & 0.59 & 0.01 & 0.08 \\
468 & --2 & 0.30 & 21.36 & 0.87 & 0.48 & 0.06 & 0.43 & 0.02 & 0.06 \\
470 & --5 & 1.00 & 20.05 & 0.87 & 0.96 & 0.96 & \nodata & 0.01 & 0.10 \\
473 & --2 & 0.40 & 21.88 & 0.82 & 0.52 & 0.26 & 0.40 & 0.02 & 0.06 \\
481 & --2 & 0.40 & 22.04 & 0.81 & 0.54 & 0.39 & 0.36 & 0.02 & 0.06 \\
482 & --5 & 0.60 & 22.16 & 0.86 & 0.23 & 0.14 & 0.21 & 0.01 & 0.04 \\
492 & 1 & 0.36 & 21.37 & 0.94 & 0.90 & 0.31 & 0.73 & 0.05 & 0.14 \\
493 & --4 & 0.78 & 22.50 & 0.81 & 0.26 & 0.18 & 0.35 & 0.01 & 0.08 \\
498 & --2 & 0.14 & 22.04 & 0.82 & 0.59 & 0.05 & 0.41 & 0.02 & 0.07 \\
507 & 2 & 0.06 & 20.91 & 0.68 & 1.17 & 0.08 & 0.74 & 0.07 & 0.11 \\
510 & --2 & 0.51 & 21.35 & 0.86 & 0.47 & 0.25 & 0.41 & 0.02 & 0.05 \\
514 & --2 & 0.72 & 21.89 & 0.80 & 0.47 & 0.34 & 0.47 & 0.03 & 0.07 \\
519 & --2 & 0.10 & 22.19 & 0.52 & 0.60 & 0.20 & 0.38 & 0.05 & 0.06 \\
523 & 0 & 0.73 & 20.83 & 0.91 & 0.99 & 0.75 & 0.91 & 0.01 & 0.05 \\
525 & 0 & 0.68 & 21.17 & 0.85 & 0.62 & 0.41 & 0.63 & 0.03 & 0.14 \\
528 & 15 & 0.00 & 21.23 & 0.51 & 1.24 & \nodata & 0.74 & 0.09 & 0.18 \\
534 & --5 & 0.44 & 21.58 & 0.89 & 0.40 & 0.14 & 0.37 & 0.01 & 0.06 \\
536 & --5 & 0.65 & 19.84 & 0.87 & 1.21 & 0.72 & 1.30 & 0.02 & 0.05 \\
537 & --2 & 0.82 & 21.46 & 0.89 & 0.31 & 0.27 & 0.26 & 0.02 & 0.07 \\
539 & --2 & 0.81 & 22.24 & 0.82 & 0.23 & 0.16 & 0.46 & 0.02 & 0.08 \\
540 & --2 & 0.54 & 21.70 & 0.89 & 0.35 & 0.12 & 0.49 & 0.02 & 0.06 \\
542 & --5 & 0.53 & 20.98 & 0.88 & 0.45 & 0.20 & 0.47 & 0.02 & 0.04 \\
544 & --2 & 0.42 & 21.16 & 0.88 & 0.73 & 0.23 & 0.69 & 0.03 & 0.13 \\
546 & --4 & 0.56 & 21.82 & 0.80 & 0.28 & 0.10 & 0.40 & 0.03 & 0.07 \\
549 & 2 & 0.17 & 21.09 & 0.76 & 0.94 & 0.15 & 0.66 & 0.02 & 0.11 \\
553 & 5 & 0.00 & 22.27 & 0.73 & 0.51 & \nodata & 0.31 & 0.03 & 0.24 \\
554 & --5 & 0.71 & 20.55 & 0.90 & 0.66 & 0.39 & 0.94 & 0.02 & 0.15 \\
555 & --2 & 0.30 & 22.35 & 0.87 & 0.37 & 0.08 & 0.31 & 0.02 & 0.05 \\
560 & 0 & 0.55 & 21.01 & 0.85 & 0.82 & 0.37 & 0.89 & 0.02 & 0.07 \\
565 & --2 & 0.74 & 22.50 & 0.68 & 0.44 & 0.48 & 0.23 & 0.00 & 0.11 \\
572 & --2 & 0.11 & 22.43 & 0.88 & 0.73 & 0.98 & 0.42 & 0.04 & 0.07 \\
584 & --5 & 0.65 & 22.18 & 0.84 & 0.37 & 0.18 & 0.54 & 0.01 & 0.10 \\
587 & 1 & 0.07 & 22.30 & 0.79 & 0.84 & 0.41 & 0.52 & 0.02 & 0.07 \\
591 & 99 & 0.46 & 22.24 & 0.55 & 1.17 & 2.01 & 0.55 & 0.08 & 0.20 \\
594 & --99 & 0.60 & 22.64 & 0.71 & 0.13 & 0.13 & 0.08 & 0.02 & 0.05 \\
626 & 1 & 0.25 & 20.94 & 0.88 & 1.03 & 0.29 & 0.76 & 0.01 & 0.07 \\
1414 & 3 & 0.65 & 22.79 & 0.78 & 0.45 & 0.75 & 0.16 & 0.01 & 0.11 \\
1455 & --99 & 0.73 & 22.78 & 0.82 & 0.33 & 0.27 & 0.26 & 0.02 & 0.08 \\
1461 & --99 & 0.40 & 23.00 & 0.77 & 0.67 & 0.59 & 0.41 & 0.01 & 0.07 \\
1475 & --2 & 0.47 & 21.82 & 0.84 & 0.61 & 0.34 & 0.48 & 0.06 & 0.09 \\
1487 & --99 & 0.00 & 24.06 & 0.46 & 0.61 & \nodata & 0.36 & 0.03 & 0.01 \\
1524 & --2 & 0.16 & 22.72 & 0.84 & 0.43 & 0.15 & 0.28 & 0.01 & 0.03 \\
1563 & --5 & 0.79 & 22.34 & 0.67 & 0.62 & 0.52 & 0.56 & 0.03 & 0.08 \\
1594 & 1 & 0.63 & 24.38 & 0.80 & 0.66 & 0.31 & 0.96 & --0.12 & 0.12 \\
1616 & --99 & 0.29 & 22.81 & 0.81 & 0.46 & 0.07 & 0.39 & 0.01 & 0.05 \\
1630 & --2 & 0.00 & 22.70 & 0.57 & 0.46 & \nodata & 0.27 & 0.03 & 0.09 \\
1649 & --4 & 0.63 & 22.37 & 0.82 & 0.34 & 0.24 & 0.29 & 0.02 & 0.07 \\
1775 & --4 & 0.36 & 22.38 & 0.80 & 0.83 & 0.58 & 0.55 & 0.04 & 0.12 \\
1806 & --2 & 0.62 & 22.59 & 0.83 & 0.54 & 0.55 & 0.32 & 0.01 & 0.08 \\
1816 & --99 & 0.69 & 23.03 & 0.80 & 0.48 & 0.47 & 0.29 & 0.01 & 0.11 \\
1829 & --99 & 0.12 & 24.40 & 0.43 & 0.44 & 0.67 & 0.26 & 0.05 & 0.18 \\
1842 & --4 & 0.11 & 23.42 & 0.73 & 1.06 & 0.93 & 0.64 & 0.01 & 0.18 \\
1865 & --5 & 0.69 & 22.22 & 0.80 & 0.54 & 0.43 & 0.43 & 0.01 & 0.11 \\
1871 & --5 & 0.71 & 21.02 & 0.89 & 0.77 & 0.47 & 1.04 & 0.04 & 0.07 \\
1897 & 999 & 0.31 & 22.34 & 0.87 & 0.76 & 0.67 & 0.47 & 0.01 & 0.11 \\
1978 & --99 & 0.15 & 23.23 & 0.90 & 0.48 & 0.24 & 0.31 & 0.03 & 0.10 \\
\enddata
\tablenotetext{a}{Hubble types from FFvD00.  Here, morphological types
are represented as E (-5), S0 (-2), Sa (1), Sb (3), Sc(5), Sd (7), Sm
(9), and Im (10); mergers are indicated with (99), conflicted types as
999, and non-typed galaxies ($m_{814}>22$) with (-99).}
\tablenotetext{b}{Measured in the F814W filter using a de Vaucouleurs
bulge with exponential disk profile.}
\tablenotetext{c}{Converted to rest-frame $B$ and $V$ using transforms
from vD98.}
\label{cl1358data}
\end{deluxetable}

\begin{deluxetable}{lrrr}
\tablecolumns{4}
\tablewidth{0pc}
\tablecaption{Average [OII]$\lambda3727$~\AA~Equivalent Widths \label{OIItab}}
\tablehead{
\colhead{$B/T$}    &    \colhead{N}   &
\colhead{$\overline{[OII]\lambda3727}$ EQ (\AA)} &       \colhead{$m_{814}$}}
\startdata
$B/T<0.25$              & 26    & $-11.7\pm1.1$ & $18.3\leq m\leq 20.9$\\
$0.25\leq B/T<0.50$     & 37    & $-1.6\pm0.9$  & $17.4\leq m\leq 21.0$\\
$0.5\leq B/T<0.75$      & 60    & $-0.3\pm0.8$  & $17.6\leq m\leq 21.0$\\
$B/T\geq0.75$           & 32    & $0.1\pm0.6$   & $17.8\leq m\leq 21.0$\\
\enddata
\end{deluxetable}

\end{document}